\pdfoutput=1 
\documentclass[12pt]{iopart}	
\usepackage{setstack, times}
\usepackage{graphicx,amssymb,amsthm,epsfig,amsbsy,amsfonts,mathrsfs}
\usepackage[english,francais]{babel}

\usepackage{url,hyperref}

\usepackage{color}

\definecolor{M_Green}          {rgb}{0.2 , 0.6 , 0.2}

\renewcommand{\leq}{\leqslant}
\renewcommand{\geq}{\geqslant}

\def\eqdef{\stackrel{\mbox{\tiny def}}{=}}     
\newcommand{\ket}[1]{|\kern.3ex#1\kern.3ex\rangle}
\newcommand{\bra}[1]{\langle\kern.3ex #1 \kern.3ex|}
\newcommand{\mean}[1]{\left\langle #1\right\rangle}
\newcommand{\smean}[1]{\langle #1\rangle}

\newcommand{\EXP}[1]{\mathrm{e}^{#1}}         

\renewcommand{\tr}[1]{\mathop{\mathrm{tr}}\nolimits\left\{ #1 \right\}}  

\newcommand{\heaviside}{\theta_\mathrm{H}}

\def\I{{\rm i}}                  
\def\D{{\rm d}}                  

\newcommand{\derivp}[2]{\frac{\partial #1}{\partial #2}}

\newcommand\identity{\mathbf{1}}

\newcommand\antiddots{\mathinner{\mkern2mu\raise1pt\hbox{.}\mkern2mu
    \newline \raise4pt\hbox{.}\mkern2mu\raise7pt\hbox{.}\mkern1mu}}

\def\Nc{N}           
\def\Nint{N_\mathrm{int}}         
\def\Sm{\mathcal{S}}       
\def\Sbar{\overline{\Sm}}  
\def\WSm{\mathcal{Q}}      
\def\Wmat{\mathcal{W}}     
\def\Kmat{\mathcal{K}}     
\def\Heff{\mathcal{H}_\mathrm{eff}}     

\def\Wt{\tau_\mathrm{W}}
\def\Ht{\tau_\mathrm{H}}
\def\invQ{\Gamma}      
\def\einvQ{\gamma}     
\def\bG{\overline{\Gamma}}

\def\rt{t}  
\def\coupl{T}
\def\fss{\kappa} 

\newcommand{\mpart}[2]{\tilde{w}_{#1,#2}} 
\newcommand{\mprop}[2]{w_{#1,#2}} 
\newcommand{\dwt}[2]{\mathscr{P}_{#1,#2}} 
\newcommand{\Rmpart}[2]{\tilde{q}_{#1,#2}} 
\newcommand{\Rmprop}[2]{q_{#1,#2}} 
\newcommand{\Rdwt}[2]{\mathscr{Q}_{#1,#2}} 

\def\tup{\tau_\mathrm{upper}} 
\def\tlow{\tau_\mathrm{lower}} 

\begin{document}


\selectlanguage{english}

\title{Wigner-Smith time-delay matrix in chaotic cavities with non-ideal contacts}

\author{Aur\'elien Grabsch}
\address{LPTMS, CNRS, Univ. Paris-Sud, Universit\'e Paris-Saclay, 91405 Orsay cedex, France}

\author{Dmitry V Savin}
\address{Department of Mathematics, Brunel University London, Uxbridge, UB8 3PH, United Kingdom}

\author{Christophe Texier}
\address{LPTMS, CNRS, Univ. Paris-Sud, Universit\'e Paris-Saclay, 91405 Orsay cedex, France}




%
%
%
%

\begin{abstract}
We consider wave propagation in a complex structure coupled to a finite number $N$ of scattering channels, such as chaotic cavities or quantum dots with external leads. Temporal aspects of the scattering process are analysed through the concept of time delays, related to the energy (or frequency) derivative of the scattering matrix $\mathcal{S}$. We develop a random matrix approach to study the statistical properties of the symmetrised Wigner-Smith time-delay matrix
$\mathcal{Q}_s = -\mathrm{i}\hbar\,\mathcal{S}^{-1/2}\big(\partial_\varepsilon\mathcal{S}\big)\,\mathcal{S}^{-1/2}$, and obtain the joint distribution of $\mathcal{S}$ and $\mathcal{Q}_s$ for the system with non-ideal contacts, characterised by a finite transmission probability (per channel) $0<T\leq1$. We derive two representations of the distribution of $\mathcal{Q}_s$ in terms of matrix integrals specified by the Dyson symmetry index $\beta=1,\,2,\,4$ (the general case of unequally coupled channels is also discussed). We apply this to the Wigner time delay $\tau_\mathrm{W}=(1/N)\,\mathrm{tr}\big\{\mathcal{Q}_s\big\}$, which is an important quantity providing the density of states of the open system. Using the obtained results, we determine the distribution $\mathscr{P}_{N,\beta}(\tau)$ of the Wigner time delay in the weak coupling limit $NT\ll1$ and identify the following three regimes.
(i)
The large deviations at small times (measured in units of the Heisenberg time) are characterised by the limiting behaviour
$\mathscr{P}_{N,\beta}(\tau)\sim\tau^{-\beta N^2/2-3/2}\,\exp\big\{-\beta N T/(8\tau)\big\}$ for $\tau\lesssim T$.
(ii)
The distribution shows the universal $\tau^{-3/2}$ behaviour in some intermediate range $T\lesssim\tau\lesssim1/(TN^2)$.
(iii)
It has a power law decay $\mathscr{P}_{N,\beta}(\tau)\sim T^2N^3(TN^2\tau)^{-2-\beta N/2}$ for large $\tau\gtrsim1/(TN^2)$.
\\[3ex]
Keywords: random matrix theory, scattering theory, quantum chaos, delay times
\\[5ex] \mbox{\ }\hfill
Published 5 September 2018 in 
\href{}{\emph{J. Phys. A: Math. Theor. \textbf{51} (2018) 404001}}
\end{abstract}

\maketitle

\renewcommand{\labelitemi}{$\bullet$}
\renewcommand{\labelitemii}{$\star$}



\section{Introduction}

Scattering of waves in complex systems has been a subject of intensive studies, with motivations ranging from compound-nucleus reactions \cite{VerWeiZir85,MitRicWei10}, coherent electronic transport \cite{Bee97} to propagation of electromagnetic waves in random media \cite{SebGen98} and chaotic billiards \cite{GuhMulWei98}. In a scattering setting, the central object is the on-shell scattering matrix $\Sm(\varepsilon)$ whose matrix elements provide the probability amplitudes of transitions (reflection and transmission) between scattering channels open at the given energy $\varepsilon$ \cite{MelKum04}. The total number $N$ of open channels is typically finite (for example, $N$ is fixed by transverse quantisation for the modes propagating in the electronic wave guides attached to a quantum dot). For an energy and flux conserving system (i.e. without losses or gain), the $\Sm$-matrix is unitary and can therefore be diagonalised as follows
\begin{equation}
\label{eq:S}
  \Sm(\varepsilon)
  =\mathcal{U}(\varepsilon)\,\EXP{\I\Theta(\varepsilon)}\,\mathcal{U}^\dagger(\varepsilon)
  \:,
  \hspace{1cm} 
  \Theta=\mathrm{diag}(\theta_1,\cdots,\theta_\Nc)\,.
\end{equation}
The diagonal matrix $\Theta$ gathers the scattering phase shifts (eigenphases) and $\mathcal{U}$ is a $\Nc\times\Nc$ unitary matrix of the corresponding eigenvectors (associated with a specific basis of solutions of the wave equation known as the \textit{partial} scattering waves). For systems invariant under time reversal, the reciprocity principle dictates $\Sm$ to be also symmetric, implying that $\mathcal{U}$ becomes an orthogonal matrix in this case.

Complementary to such a stationary description, the temporal aspects of the scattering process may also be characterised in terms of the $\Sm$-matrix by several means. The most well-known concept is probably that of resonance widths, which are related to finite lifetimes of resonance states formed at the intermediate stage of the scattering event \cite{Kot05,FyoSav11}. Such resonances are formally defined through the analytical structure of $\Sm(\varepsilon)$ in the complex $\varepsilon$ plane, corresponding to the poles $\mathcal{E}_n=E_n-\frac{\I}{2}\Gamma_n$, where $E_n$ and $\Gamma_n>0$ are the energy and width of the $n$th resonance, respectively. Practically, they are accessible by performing the spectroscopy analysis of relevant decay spectra \cite{GluKolKor02,KuhHohMaiSto08,DiFKraFra12}.

The \textit{time delay} is another important notion used to quantify the duration of the scattering event. Following Wigner \cite{Wig55} and Smith \cite{Smi60}, the time spent by an incident wave in the scattering region can be characterized in terms of the following matrix:
\begin{equation}
\label{eq:Q_WS}
  \WSm(\varepsilon) = -\I\hbar \Sm^\dagger(\varepsilon) \derivp{\Sm(\varepsilon)}{\varepsilon}.
\end{equation}
Below we set $\hbar=1$. The Wigner-Smith matrix (\ref{eq:Q_WS}) is Hermitian by construction (for unitary $\Sm$) and thus has all real eigenvalues $\{\tau_1,\cdots,\tau_\Nc\}$ that are commonly referred to as \textit{proper time delays}. They provide the lifetimes of metastable states. On the other hand, the diagonal elements $\{\WSm_{11},\cdots,\WSm_{\Nc\Nc}\}$ of (\ref{eq:Q_WS}) are also real and serve to characterise the time delay in a given entrance channel \cite{Smi60,Lyu77}. Taking the trace, one arrives at the simple averaged characteristic, the so-called Wigner time delay \cite{Lyu77,LewWei91,LehSavSokSom95,FyoSom97}
\begin{equation}
\label{eq:Wig_def}
  \Wt(\varepsilon)  \equiv\frac{1}{\Nc}\tr{\WSm}
  = -\frac{\I}{\Nc} \derivp{\ln\det\Sm(\varepsilon)}{\varepsilon}
  \,.
\end{equation}
This quantity plays an important role in practical applications \cite{CarNus02,KolVos13,Tex16}. In particular, it provides a measure of the density of states of the open system, thus being essential for the description of electronic and transport properties of coherent conductors~\cite{Tex16}.

In view of representations (\ref{eq:S}) and (\ref{eq:Wig_def}), it is also instructive to consider the so-called \textit{partial time delays}
$\{\tilde{\tau}_a =  \partial\theta_a/\partial\varepsilon$, $a=1,\cdots,\Nc\}$ that are defined by the energy derivative of the scattering eigenphases \cite{FyoSom97}. They may be treated as the time delay of a ``narrow'' wave packet (with a weak energy dispersion around $\varepsilon$) prepared with respect to a given scattering eigenchannel. Although there is a connection between those three time delay sets \cite{SavFyoSom01}, they generally characterise different aspects of the problem~\cite{Tex16}. Other characteristic times can also be introduced using certain derivatives of the $\Sm$-matrix elements, see reviews \cite{CarNus02,KolVos13,Tex16,HauSto89,But90,LanMar94,But02a} for relevant studies.

Generally, time delays are known to satisfy certain inequalities (essentially imposed by causality); in particular, they cannot take arbitrary large negative values \cite{CarNus02,KolVos13}. In the resonance approximation, however, one can neglect a smooth energy dependence associated with potential scattering and direct reactions. The whole dependence of $\Sm(\varepsilon)$ on the energy is then due to its complex poles (resonances). Under such assumptions, the Wigner-Smith matrix becomes strictly positive \cite{SokZel97}. The Wigner time delay is then determined entirely by the resonance spectrum of the open system \cite{Lyu77,LehSavSokSom95}
\begin{equation}\label{eq:Wig_res}
  \Wt(\varepsilon)
  = \frac{1}{\Nc}\sum_n\frac{\Gamma_n}{(\varepsilon-E_n)^2+\Gamma_n^2/4}\,.
\end{equation}
This important expression is valid at arbitrary degree of the resonance overlap, leading to an interpretation of the Wigner time delay (\ref{eq:Wig_res}) as the density of states in open systems, see \cite{FyoSom97,CarNus02,Tex16} for further discussion. The spectral average of $\Wt$ over a narrow energy window is given by $\overline{\tau}_\mathrm{W} = 2\pi/(N\Delta)$, where $\Delta$ is the mean level spacing (which carries a smooth $\varepsilon$-dependence in general). This relates the mean time delay $\overline{\tau}_\mathrm{W}=\Ht/N$ to the fundamental timescale of quantum systems, the Heisenberg time $\Ht=2\pi/\Delta$.

When these concepts are applied to complex quantum or wave systems, such as quantum dots or microwave billiards with classically chaotic dynamics, a statistical analysis is required in order to characterise strong fluctuations that arise in scattering. There are two main approaches to describe such fluctuations: the semiclassical method (see \cite{KuiSavSie14} for recent advances and further references) and random matrix theory (RMT). The latter proved to be extremely successful in describing universal patterns of chaotic wave phenomena \cite{GuhMulWei98}, being also the most suitable in order to provide the full statistical information in terms of both correlations and distributions.

There are two possible variants of the RMT formulation of chaotic scattering.
The \textit{stochastic approach} (see \cite{Bee97,MelKum04} for reviews) treats the $\Sm$-matrix as the prime statistical object without any reference to the system Hamiltonian. The probability distribution of $\Sm=\Sm(\varepsilon)$ (at the fixed energy $\varepsilon$) is deduced from a maximum entropy principle subject to the global constraints imposed on $\Sm$ by its symmetry and analyticity. It is given by the so-called Poisson kernel~\cite{MelPerSel85,Bro95,MelKum04}
\begin{equation}
  \label{eq:PoissonKernel}
  P_{\Sm}(\Sm)\propto|\det(\identity_\Nc-\Sbar^*\Sm)|^{-2-\beta(\Nc-1)}\,,
\end{equation}
which is parameterised by the mean (``optical'') scattering matrix $\Sbar$. In the absence of direct reactions,  $\Sbar$ can always be chosen as a constant diagonal matrix \cite{EngWei73}. The symmetry index $\beta=1$ ($\beta=2$) corresponds to the systems with preserved (broken) time-reversal symmetry ($\beta=4$ is to be taken when spin rotational symmetry is broken). The approach proved to be very useful, in particular, for studying coherent electronic transport in mesoscopic systems \cite{Bee97}. However, correlations at different energies as well as other spectral properties of open systems related to the resonances turn out to be inaccessible in such an approach because of its fixed-energy nature (in this respect, see \cite{BroBut97} for an extension to address the energy dependence).

The \textit{Hamiltonian approach} \cite{MahWei69,VerWeiZir85} is the other and more general formulation that is well adapted to treat both scattering and spectral characteristics on equal footing \cite{SokZel89,FyoSav11}. Within the resonance approximation considered, the starting point is the following representation of the $\Sm$-matrix in terms of the Wigner's reaction matrix $\Kmat$:
\begin{equation}
 \label{eq:SandK}
  \Sm(\varepsilon) = \frac{\identity_\Nc-\I\,\Kmat(\varepsilon)}{\identity_\Nc+\I\,\Kmat(\varepsilon)}\,,
  \qquad
  \Kmat(\varepsilon) = \Wmat^\dagger (\varepsilon-\mathcal{H})^{-1}\Wmat\,.
\end{equation}
Here, the Hermitian matrix $\mathcal{H}$ of size $\Nint$ represents the internal Hamiltonian of the closed system, whereas the rectangular $\Nint\times\Nc$ matrix $\Wmat$ consists of the constant coupling amplitudes between $\Nint$ internal and $\Nc$ channel states. In the chaotic regime, $\mathcal{H}$ is modelled by an RMT ensemble of appropriate symmetry \cite{GuhMulWei98}. In the RMT limit $\Nint\gg1$, spectral fluctuations become universal (model-independent) on the local scale of the mean level spacing $\Delta$. Similarly, the results turn out to be insensitive to particular statistical assumptions on the amplitudes $\{\Wmat_{na}\}$ provided that $\Nc\ll\Nint$ \cite{LehSavSokSom95,LehSahSokSom95}. These amplitudes appear in the final expressions only through the transmission coefficients
\begin{equation}
  \label{eq:RelationBetweenCouplingParameters}
 \coupl_a \equiv 1-|\Sbar_{aa}|^2 = \frac{4\fss_a}{(1+\fss_a)^2}\,,
 \qquad
 \fss_a = \frac{2\pi\|\Wmat_a\|^2}{\Nint\Delta}\,.
\end{equation}
$\coupl_a$ describes the probability of entering the system through channel $a$ (thus characterizing the contact quality), with $\coupl_a\ll1$ ($\coupl_a=1$) corresponding to weak (perfect) coupling.

 The Hamiltonian approach, especially when combined with the supersymmetry technique to perform statistical averages \cite{VerWeiZir85}, offers the powerful method to derive exact non-perturbative results for various correlation and distribution functions at any channel coupling, see \cite{MitRicWei10,FyoSav11,FyoSavSom05} for details. It was actually possible to derive the Poisson kernel (\ref{eq:PoissonKernel}) starting from representation (\ref{eq:SandK}), thus proving  equivalence of the two approaches for the $\Sm$-matrix distribution \cite{Bro95}. As to the Wigner-Smith matrix, a number of exact results are already known for various time delays at any $\coupl_a\leq1$, which we will briefly overview below. However, the distribution of the whole $\WSm$ matrix (in its symmetrised form) is only known for the special case of perfect coupling (all $T_a=1$) \cite{BroFraBee97,BroFraBee99}. It is the aim of this work to fill in this gap and to provide the corresponding distribution at arbitrary coupling. We derive the exact result in terms of certain matrix integrals and further analyse the relevant marginal densities in the weak coupling limit.

The outline of the paper is as follows. In the next section we state the main results of this work. In Section~\ref{sec:Motivations}, we first provide a heuristic analysis of the Wigner time delay distribution  in the weak coupling limit, providing some physical intuition on the nature of the results; then we give an overview of the known exact results at arbitrary $\coupl$ (this overview is complemented by \ref{app:PartialProper}). Section~\ref{sec:model} develops a resonance representation of the Wigner-Smith matrix. The mapping between the perfect and arbitrary coupling is established and used in Section~\ref{sec:WSMdistribution} to derive a general representation for the Wigner-Smith matrix distribution at arbitrary coupling in terms of a matrix integral over Hermitian matrices. In Section~\ref{sec:IntegrationOverCUE}, we work out an alternative form of the distribution in terms of a matrix integral over the unitary group, which turns out to be more useful for numerics. Based on these results, we study the characteristic function of the Wigner time delay in the weak coupling limit in Section~\ref{sec:characfct} and deduce the limiting behaviours of its distribution. Some numerical analysis is presented in Section~\ref{sec:Numerics}. Finally, we provide several appendices with more technical details of our calculations, which we believe may be helpful for further developments and applications.


\section{Statement of the main results}

We consider the symmetrised Wigner-Smith matrix defined by
\begin{equation}\label{eq:WS_sym}
  \WSm_{s} = \Sm^{1/2}\WSm\Sm^{-1/2} = -\I\,\Sm^{-1/2} \partial_\varepsilon\Sm\,\Sm^{-1/2}\,,
\end{equation}
which clearly has the same spectrum as $\WSm$. Our first main result is the joint matrix distribution for the scattering matrix $\Sm$ and the inverse matrix $\invQ=\WSm_s^{-1}$ at arbitrary transmission of $\Nc$ channels. To this end, we develop a resonance representation for the Wigner-Smith matrix to establish a relation between this matrix at arbitrary and perfect coupling. This enables us to apply the known joint distribution at perfect coupling \cite{BroFraBee97,BroFraBee99} to that at arbitrary coupling. When all channels have the same transmission coefficient $T=1-|\Sbar|^2$, our result reads
\begin{eqnarray}
  \label{eq:JointMatrixDistribIntro}
  &&\hspace{-1cm}
  \mathrm{D}\Sm\,\,\mathrm{D}\invQ\, P_{\Sm,\invQ}(\Sm,\invQ)
  =
  \mathrm{D}\Sm\,\,\mathrm{D}\invQ\,
  c_{\Nc,\beta}  \,
  \Theta(\invQ)\,
  \left|\det\big(\identity_\Nc-\Sbar^*\Sm\big)\right|^{\beta\Nc}\,
  \\\nonumber
  &&\times
  \big(\det\invQ\big)^{\beta\Nc/2} \,
  \exp\left[
     -\frac{\beta}{2(1-|\Sbar|^2)}\,
     \tr{(\identity_\Nc-\Sbar^*\Sm)(\identity_\Nc-\Sbar\Sm^\dagger)\invQ}
  \right],
\end{eqnarray}
where $c_{\Nc,\beta}$ is a normalisation constant. $\mathrm{D}\Sm$ is the Haar measure (uniform measure over unitary matrices) and $\mathrm{D}\invQ$ the Lebesgue measure over the set of Hermitian matrices. The matrix theta function is $\Theta(\invQ)=1$ when all eigenvalues of $\invQ$ are positive and zero otherwise.
The result \eref{eq:JointMatrixDistribIntro} has relied on the following conservation of the measure when mapping the $\Sm$ and $\invQ$  matrices for the ideal and non-ideal contacts:
\begin{equation*} \nonumber
  \mathrm{D}\Sm_0\,\,\mathrm{D}\invQ_0=\mathrm{D}\Sm\,\,\mathrm{D}\invQ
  \:.
\end{equation*}

The representation \eref{eq:JointMatrixDistribIntro} may be regarded as an extension of the Poisson kernel \eref{eq:PoissonKernel} for the time-delay problem.
One can then deduce the distribution of the matrix $\invQ$ in terms of a matrix integral over the unitary group.
We have prefered a more convenient form, induced by \eref{eq:SandK}, in terms of a matrix integral over Hermitian matrices
\begin{eqnarray}
  \label{eq:DistribInvQintro}
  &&\hspace{-1cm}
  P_\invQ(\invQ) = b_{\Nc,\beta}
    \Theta(\invQ)\,
    ( \det\invQ )^{\beta\Nc/2}
  \\\nonumber
  &&\times
    \int\mathrm{D}\Kmat\,
    \frac{\det(\identity_\Nc+\Kmat^2)^{\beta\Nc/2}}
         {\det(\identity_\Nc+\fss^2\Kmat^2)^{1-\frac{\beta}{2}+\beta\Nc}}\,
    \exp\left[
      -\frac{\beta}{2}\fss\,\tr{ \frac{\identity_\Nc+\Kmat^2}{\identity_\Nc+\fss^2\Kmat^2}\invQ }
    \right]
\end{eqnarray}
where $b_{\Nc,\beta}$ is a normalisation constant  and the coupling constant $\fss>0$ is related to the transmission coefficient \eref{eq:RelationBetweenCouplingParameters}. For $\fss=1$ ($\Sbar=0$), Eq.~\eref{eq:DistribInvQintro} reduces to the Laguerre ensemble corresponding to the known result at prefect coupling \cite{BroFraBee97,BroFraBee99}. We have also generalised this expression to the most general case when channels are not equally coupled, see equation~\eref{eq:DistribGammaArbitraryCouplings} below in the text.

The matrix distribution \eref{eq:DistribInvQintro} is further used to study the distribution $\dwt{\Nc}{\beta}(\tau)$ of the Wigner time delay $\Wt=(1/\Nc)\tr{\invQ^{-1}}$. Defining the characteristic function (Laplace transform of the distribution) as
$
  \mathcal{Z}_{\Nc,\beta}(p)\propto  \mean{ \exp\big\{ -(2p/\beta\fss)\tr{\invQ^{-1}} \big\}  }
$,
which involves in principle two matrix integrals (over $\invQ$ and $\Kmat$), we have finally obtained a ratio of two $\Nc\times\Nc$ determinants integrated over the eigenvalues of one matrix only
\begin{eqnarray}
  \label{eq:Zn2intro}
  \hspace{-1.5cm}
  \mathcal{Z}_{\Nc,2}(p)
  =\int_{\mathbb{R}^\Nc}
  \frac{ \D k_1\cdots\D k_\Nc\, \Delta_\Nc(k)^2}{ \prod_n(1+k_n^2)^{\Nc} }
  \,
  \frac{
  \det\left[
    \left(
       p\,
       \frac{1+\fss^2k_j^2}{1+k_j^2}
    \right)^{\frac{\Nc+i}{2}}
    K_{\Nc+i} \left(2\sqrt{p\frac{1+k_j^2}{1+\fss^2k_j^2}}\right)
  \right]
  }{
  \det\left[
    \left(
       \frac{1+k_j^2}{1+\fss^2k_j^2}
    \right)^{-\Nc-i}
  \right]
  }
  \:.
\end{eqnarray}
The result holds in the unitary ($\beta=2$) case. Here $\Delta_\Nc(k)=\prod_{i<j}(k_i-k_j)$ denotes the Vandermonde determinant and $K_\nu(x)$ is the MacDonald function (modified Bessel function of 3rd kind). Because \eref{eq:Zn2intro} has a finite limit when $\fss\to0$, this result shows, in particular, that when rescaled by the factor $\fss\simeq\coupl/4\to0$, the Wigner time delay distribution has a limit
$\lim_{\fss\to0}\fss\,\dwt{\Nc}{2}(\fss\,\rt)=\Rdwt{\Nc}{2}(\rt)$ independent of~$\fss$. We have also verified numerically that this holds for all $\beta$.

Finally we have obtained the limiting behaviours of the distribution $\dwt{\Nc}{\beta}(\tau)$ in the weak coupling limit, $\Nc\coupl\ll1$. The large deviations $\tau\to0$ are characterised by
\begin{equation}
  \dwt{\Nc}{\beta}(\tau)
  \sim
  \tau^{-\frac{\beta\Nc^2}{2}-\frac32}\,
  \EXP{-\beta\Nc\coupl/(8\tau)}
  \hspace{1cm}
  \mbox{for } \tau\ll \coupl\,,
\end{equation}
which is obtained by extending the steepest descent method to the matrix integrals (for arbitrary symmetry class).
For $\beta=2$, we have also deduced this behaviour from \eref{eq:Zn2intro}.
Then, analysing in detail the limit first $\fss\to0$ and then $p\to0$ of the characteristic function \eref{eq:Zn2intro}, we have obtained the power law
\begin{equation}
  \dwt{\Nc}{\beta}(\tau)
  \sim
  \frac{1}{\coupl}(\coupl/\tau)^{3/2}
  \hspace{1cm}
  \mbox{for }
  \coupl \ll \tau \ll 1/(\Nc^2\coupl)
  \:,
\end{equation}
which holds independently of $\beta$. Finally, we have provided a simple argument in terms of isolated resonances to get the large time asymptotic as follows
\begin{equation}
  \dwt{\Nc}{\beta}(\tau)
  \sim
  \coupl^2\Nc^3
  \,
  \left( \coupl\Nc^2\,\tau\right)^{-2-\beta\Nc/2}
\end{equation}
All these limiting behaviours have been verified numerically.


\section{Background and motivations}
\label{sec:Motivations}

\subsection{Heuristic analysis (single resonance approximation)}
\label{Subsec:Heuristic}

Before entering into the detailed analysis, it is instructive to give a qualitative discussion of the typical behaviour of time delay distributions when all $\coupl_a=\coupl\ll1$. The channel coupling can be treated perturbatively in such a case, enabling us to estimate the mean width (decay rate) by the Fermi's golden rule as $\bG=\Nc\coupl\frac{\Delta}{2\pi}=\Nc\coupl/\Ht$. (This is known as the Weisskopf estimate in nuclear physics and as the inverse of the dwell time in mesoscopics.) The distribution of the resonance widths (rescaled in units of $\bG$) is then given by the well known $\chi^2$ distribution with $\beta\Nc$ degrees of freedom,
\begin{equation}
  \label{eq:DistribResonanceWidth}
  p (y) =
  \frac{ (\beta\Nc/2)^{\beta\Nc/2} }{\Gamma(\beta\Nc/2)} \,
  y^{\frac{\beta\Nc}{2}-1}\EXP{-\beta\Nc y /2}\,,
  \qquad y\equiv\Gamma/\bG=\Gamma\Ht/(\Nc\coupl)\,.
\end{equation}
The related moments are
$\smean{y^k} = \big(\frac{2}{\beta\Nc}\big)^{k} \big(\frac{\beta\Nc}{2}\big)_k$, where $\mean{\cdots}$ denotes the statistical averaging and $(a)_k$ is the Pochhammer symbol. Expression \eref{eq:DistribResonanceWidth} is a many-channel generalisation of the famous Porter-Thomas result at $\Nc=1$ and $\beta=1$ \cite{PorTho56}. It must be emphasised that this distribution arises from a perturbative treatment of the channel coupling, resulting essentially from Gaussian statistics of the chaotic wave functions of the closed system. Thus it is valid in the \textit{weak} coupling limit only. (Notably at perfect coupling, the exact distribution of resonance widths is known \cite{FyoSom97} to develop the power law decay $p(y)\propto y^{-2}$ at $y\gg1$. See Ref.~\cite{FyoSav15} for further discussion of the weak coupling limit beyond the perturbative regime.)

We now consider the important case of \emph{isolated} (well-separated) resonances when typical widths $\Gamma_n\ll|E_n-E_{n+1}|\sim\Delta$, corresponding to $2\pi\bG/\Delta=\Nc\coupl\ll1$. Scattering patterns in such a regime are dominated by a single resonance with energy $E_n\approx\varepsilon$ closest to the scattering energy. Accordingly, we may approximate the Wigner time delay (\ref{eq:Wig_res}) as $\Wt(\varepsilon)\simeq(1/\Nc)\,\Gamma_n/\big[(\varepsilon-E_n)^2+\Gamma_n^2/4\big]$ and assume the statistically uncorrelated energies and widths \cite{FyoSom97}. This leads to the following  form of the time delay distribution:
\begin{equation}
  \label{eq:heuristic0}
  \dwt{\Nc}{\beta}\left( \tau \right)
  \approx
  \int_0^\infty \frac{\D\Gamma}{\bG}\, p\left(\frac{\Gamma}{\bG}\right)
  \int_{-\Delta/2}^{\Delta/2}\frac{\D E}{\Delta}\,
  \delta\!\left(
    \tau - \frac{1}{\Nc}\,\frac{\Gamma}{E^2+\Gamma^2/4}
  \right)
  \:.
\end{equation}
Since such a Lorentzian profile is the most natural shape of the energy dependence in the vicinity of the resonance, one may generally expect that approximation (\ref{eq:heuristic0}) describes adequately the other types of time delays in the limit considered as well.

In the regime $\tau>2/(\Nc\Delta)=\Ht/(\Nc\pi)$, the cutoff of the integration over $E$ at $\Delta/2$ plays no role and can be replaced by infinity. We obtain the useful representation
\begin{equation}
  \dwt{\Nc}{\beta}\left( \tau \right)
  \approx
  \frac{1}{\Delta\sqrt{\Nc}\,\tau^{3/2}}
  \int_0^{4/(\Nc\tau)}
  \frac{\D\Gamma}{\bG}\, p\left(\frac{\Gamma}{\bG}\right)
  \frac{\sqrt{\Gamma}}{\sqrt{1-\Gamma\Nc\tau/4}}
  \:.
\end{equation}
In the intermediate regime $\Ht/\Nc\ll\tau\ll1/(\Nc\bG)=\Ht/(\Nc^2\coupl)$, the main contribution comes from the most probable (typical) resonances of width $\Gamma\sim\bG$, resulting in
\begin{equation}
  \label{eq:heuristic1}
  \dwt{\Nc}{\beta}(\tau)
  \sim
  \frac{1}{\coupl\Ht}
  \left(\frac{\coupl\,\Ht}{\tau}\right)^{3/2}
\end{equation}
independently of the symmetry index $\beta$. Such a behaviour was already observed in previous studies of both partial \cite{FyoSom97,FyoSavSom97} and proper time delays \cite{SomSavSok01}. It is believed that this $\tau^{-3/2}$ law is the most robust feature of the distribution in the regime of isolated resonances. Finally, the far tail of the distribution at $\tau\gg1/(\Nc\bG)=\Ht/(\Nc^2\coupl)$ is controlled by rare narrow resonances of width $\Gamma\ll\bG$, yielding the asymptotic behaviour
\begin{equation}
  \label{eq:heuristic2}
  \dwt{\Nc}{\beta}(\tau)
  \sim
  \frac{\coupl^2\Nc^3}{\Ht}
  \,
  \left(\frac{\Ht}{\coupl\Nc^2\,\tau}\right)^{2+\beta\Nc/2}.
\end{equation}
The universal exponent $2+\beta\Nc/2$ can be simply understood from the limiting behaviour of the resonance width distribution (\ref{eq:DistribResonanceWidth}) at $\Gamma\to0$, as explained in Ref.~\cite{FyoSom97}.

As a check, we can estimate the moments from this distribution.
The (positive) moments are controlled by the upper cutoff $\tau_*\sim\Ht/(\Nc^2\coupl)$ of the power law \eref{eq:heuristic1}. Thus we obtain
$\mean{\Wt^k}\sim\sqrt{\coupl\Ht}\,\tau_*^{k-1/2}$ i.e.
\begin{equation}
  \mean{\Wt^k} \sim \Ht^k/\big(\coupl^{k-1}\Nc^{2k-1}\big),
  \qquad\mbox{for } k<1+\beta\Nc/2
  \:,
\end{equation}
whereas all moments of higher order $k\geq1+\beta\Nc/2$ diverge because of Eq.~(\ref{eq:heuristic2}). In the $\Nc\coupl\ll1$ limit, this reproduces the known exact results (for $k=1,\,2$) discussed below.

One of our aims will be to settle these results and analysis on rigorous grounds and in particular to characterise the large deviations for $\tau\to0$.

\subsection{Known exact results}

As mentioned in the Introduction, the three time-delay sets in question do not coincide in general and have different statistical properties. We note that the formal order of the two operations, the diagonalisation and taking the energy derivative of the scattering matrix, is reversed when dealing with the proper or partial time delays. The connection between the two sets can in principle be found from the general expression \cite{SavFyoSom01}
\begin{equation}\label{eq:Q_prop-part}
  \WSm
  = \mathcal{U}\,\partial_\varepsilon\Theta\,\mathcal{U}^\dagger
  + \I\,\Sm^\dagger\left[\mathcal{U}\,\partial_\varepsilon\mathcal{U}^\dagger,\,\Sm \right],
\end{equation}
where $\partial_\varepsilon\equiv\derivp{}{\varepsilon}$ and $[\,,\,]$ stands for a commutator. This clearly shows that the differences between the proper and partial time delays are due to the second term in (\ref{eq:Q_prop-part}), which essentially accounts for the different bases chosen to express the $\Sm$-matrix~\cite{Tex16}. Clearly, the time delays satisfy the following sum rule:
\begin{equation}
  \label{eq:SumRulePPW}
  \frac{1}{\Nc}\sum_{a=1}^\Nc\tau_a
  =\frac{1}{\Nc}\sum_{a=1}^\Nc\tilde{\tau}_a
  =\frac{1}{\Nc}\sum_{a=1}^\Nc\WSm_{aa}
  =\frac{1}{\Nc} \tr{\WSm}
  = \Wt\,.
\end{equation}
In the case of the equivalent channels, this sum rule implies the following equality for the mean time delays: $\smean{\tau_a}=\smean{\tilde\tau_a}=\smean{\WSm_{aa}}=\smean{\Wt}=\frac{\Ht}{\Nc}$. It is therefore useful to measure all times in units of the Heisenberg time, and simply set $\Ht=1$ below.

\subsubsection{Perfect coupling, $T=1$.}

In the special case of one open channel, $\Nc=1$, all time delays reduce to a single quantity, the energy derivative of the scattering phase.
Its distribution was first derived for $\beta=2$ (but any $\coupl$) in Ref.~\cite{FyoSom96} and independently for any $\beta$ (but $\coupl=1$) in Ref.~\cite{GopMelBut96}.
The matrix generalisation of the latter approach to arbitrary $\Nc>1$ was presented in the influential work \cite{BroFraBee97,BroFraBee99} by Brouwer, Frahm, and Beenakker (BFB) who showed that the proper time delays (more precisely, their inverses) are distributed according to the Laguerre ensemble of random matrices. This provided a route for applying powerful RMT techniques (like orthogonal polynomials and the Coulomb gas method) to study various densities, moments and correlators built on the Wigner-Smith matrix \cite{SavFyoSom01,MezSim11,MezSim12,MezSim13,TexMaj13,MarMarGar14,GraTex15,Cun15,Nov15a,CunMezSimViv16,CunMezSimViv16b,GraMajTex17b}. We refer to Ref.~\cite{Tex16} for the most recent review and briefly discuss below the qualitative differences in the behaviour of the relevant distribution functions (more details on the marginal distributions of both proper and partial time delays can be found in~\ref{app:PartialProper}).

The many-channel distribution of the Wigner time delay is explicitly known only for $\Nc=2$~\cite{SavFyoSom01} or $\Nc\gg1$~\cite{TexMaj13}. However, its variance can be found exactly at any $\Nc$ and is represented by the following form (valid for arbitrary $\beta$ considered)~\cite{MezSim13}:
\begin{equation}
  \label{eq:VarianceWTDperfect}
 \mathrm{var}(\Wt) = \frac{4}{\Nc^2(\Nc+1)(\beta\Nc-2)} \simeq \frac{4}{\beta\Nc^4}
 \:.
\end{equation}
Here and below the symbol $\simeq$ is used to show the leading asymptotic at $\Nc\gg1$. For the partial time delays, the variances and covariances can be derived from the exactly known marginal~\cite{FyoSom96,FyoSom97,FyoSavSom97} and joint (two-point) \cite{SavFyoSom01} densities and read as follows \cite{KuiSavSie14}~:
\begin{eqnarray} \label{eq:var_part}
  \mathrm{var}(\tilde{\tau}_a) &= \frac{2}{\Nc^2(\beta\Nc-2)}\simeq\frac{2}{\beta\Nc^3}
  \:,
  \\
  \label{eq:CovPartial}
  \mathrm{cov}(\tilde{\tau}_a,\tilde{\tau}_b) &= + \frac{\mathrm{var}(\tilde{\tau}_a)}{\Nc+1}
  \simeq +\frac{2}{\beta\Nc^4}
  \simeq +\frac{1}{\Nc}\,\mathrm{var}(\tilde{\tau}_a)
  \:.
\end{eqnarray}
The corresponding expressions are also available for the proper time delays ~\cite{MezSim11,MarMarGar14},
\begin{eqnarray}
  \mathrm{var}(\tau_a) &= \frac{\Nc[\beta(\Nc-1)+2]+2}{\Nc^2(\Nc+1)(\beta\Nc-2)}\simeq\frac{1}{\Nc^2}
  \:,
  \\
  \label{eq:CovProper}
  \mathrm{cov}(\tau_a,\tau_b) &= - \frac{1}{\Nc^2(\Nc+1)}\simeq-\frac{1}{\Nc^3}
  \simeq -\frac{1}{\Nc}\,\mathrm{var}(\tilde{\tau}_a)
  \:.
\end{eqnarray}
As to the diagonal elements $\WSm_{aa}$, less is known about their statistical properties  except for the $\beta=2$ case (unitary symmetry), when it can be shown that the distributions of  $\WSm_{aa}$ and $\tilde{\tau}_a$ coincide~\cite{SavFyoSom01}. This follows from the general relation $\tilde{\tau}_a=\big[\mathcal{U}^\dagger\,\WSm\,\mathcal{U}\big]_{aa}$, implied by (\ref{eq:Q_prop-part}), and from the statistical independence of $\mathcal{U}$ and $\WSm$ in that case. For the $\beta=1$ case (orthogonal symmetry), these two matrices become statistically correlated, resulting in different statistics, in particular, $\mathrm{var}(\tilde{\tau}_a)\neq\mathrm{var}(\WSm_{aa})$. The latter variance was recently computed at any $\Nc$ in \cite{KuiSavSie14} using semiclassical methods, yielding $\mathrm{var}(\WSm_{aa})\simeq1/N^3$ unlike the $2/N^3$ dependence of (\ref{eq:var_part}) (but the exact RMT result is still lacking). It is also worth
mentioning the recent study of the distribution of $\sum_{a=1}^K\WSm_{aa}$, where the sum is restricted to a fraction of terms $K<\Nc$ (cf. Appendix of Ref.~\cite{GraMajTex17b}).

\begin{figure}[!ht]
\centering
\includegraphics[scale=0.9]{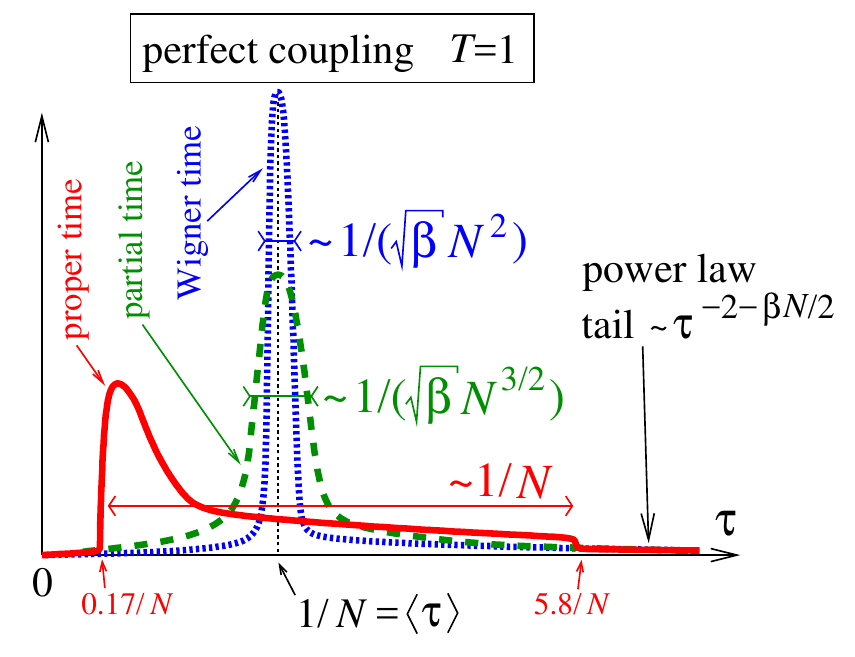}
\hfill
\includegraphics[scale=0.9]{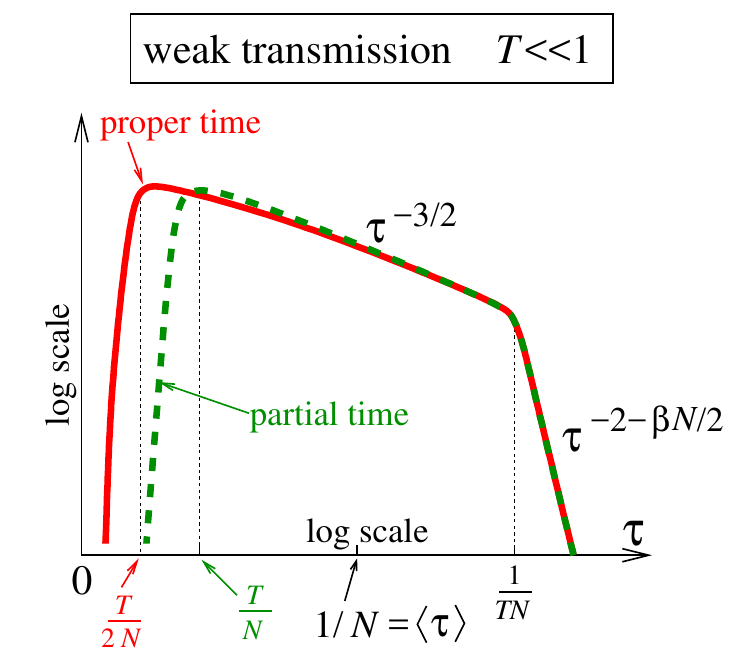}
\caption{\it
 Sketch of the time delay distributions for $\Nc\gg1$ equivalent channels at perfect (left) or weak coupling (right), where $T$ is the channel transmission coefficient. Shown are the distributions of the proper (continuous red line) and partial time delays (dashed green line) as well as the Wigner time delay distribution (dotted blue line), with all times being measured in units of the Heisenberg time.  For perfect coupling, the proper time delay distribution is close to the Mar\v{c}enko-Pastur law (with additional large deviation tails out of the interval $[(\sqrt{2}-1)^2/\Nc,(\sqrt{2}+1)^2/\Nc]$; cf.~\cite{Tex16}). For weak coupling, the distribution of the Wigner time delay is not shown, as it is still unknown and will be determined in the present paper.}
  \label{fig:SketchPC}
\end{figure}

The time delays in question clearly show the different scaling with $\Nc$, dependence in $\beta$ and sign of the correlations at perfect coupling. This leads to the profound differences between the corresponding distributions, which are schematically illustrated in Fig.~\ref{fig:SketchPC}.

\subsubsection{Non-ideal contacts, $T<1$.}

The general case of arbitrary transmission is more challenging for rigorous analysis, with studies being restricted to certain correlation functions and marginal distributions only. Most of the results have been derived within a nonperturbative approach developed in \cite{LehSavSokSom95,FyoSom97,SomSavSok01}, which can be also extended to include effects of finite absorption \cite{SavSom03} and disorder \cite{FyoSavSom05,OssFyo05}. In particular, the variance of the Wigner time delay follows from the autocorrelation function of $\Wt(\varepsilon)$ which is known exactly for both orthogonal ($\beta=1$) \cite{LehSavSokSom95} and unitary ($\beta=2$) symmetry \cite{FyoSom96} as well as in the whole crossover region between the two cases \cite{FyoSavSom97}. The variance is found to take a simple explicit form only in the $\beta=2$ case \cite{FyoSom96,FyoSom97}, being given then by
\begin{equation}
  \label{eq:FyodorovSommers1997Eq195}
   \mathrm{var}(\Wt) = \frac{2\left[1-(1-\coupl)^{\Nc+1}\right]}{\coupl^2\Nc^2(\Nc^2-1)}\,.
\end{equation}
Considering the limit of weak transmission per channel $\coupl\ll1$, we get from \eref{eq:FyodorovSommers1997Eq195} two different possible behaviours, which depend on the product $\Nc\coupl$ describing the degree of the resonance overlap (thus controlling the overall coupling to the continuum)~:
\begin{equation}  \label{eq:VarianceWT}
  \frac{ \mathrm{var}(\Wt) }{ \mean{\Wt}^2 }
  \underset{\coupl\ll1}{\simeq}
  \left\{
  \begin{array}{ll}
    \displaystyle
    \frac{4}{\beta(\Nc\coupl)^2} \ll 1 & \mbox{for } \Nc\coupl\gg1
    \\[0.5cm]
    \displaystyle
    \frac{2}{\Nc\coupl} \gg 1 & \mbox{for } \Nc\coupl\ll1
  \end{array}
  \right. .
\end{equation}
Here we have reintroduced $\beta$
to match the known $\beta=1$ result \cite{LehSavSokSom95}. The marginal distribution of the partial time delays is known exactly at any $\beta$ \cite{FyoSom97,FyoSavSom97}. The corresponding variance is also given by a simple explicit expression for $\beta=2$~\cite{FyoSom97}~:
\begin{equation}
  \label{eq:FyodorovSommers1997Eq167}
  \mathrm{var}(\tilde{\tau}_a)= \frac{2\Nc(\coupl^{-1}-1)+1}{\Nc^2(\Nc-1)}
   \underset{\coupl\ll1}{\simeq}
   \frac{2}{\coupl\,\Nc^2}
   \:.
\end{equation}
By combining \eref{eq:FyodorovSommers1997Eq195} and \eref{eq:FyodorovSommers1997Eq167}, we readily deduce an exact result for the covariance~:
\begin{eqnarray}
  \label{eq:CovPartialWeakCoupling}
  \hspace{-2cm}
  \mathrm{cov}(\tilde{\tau}_a,\tilde{\tau}_b)
  &=
  \frac{1}{\coupl^2\Nc(\Nc-1)^2}
  \left[
     \frac{2\left[1-(1-\coupl)^{\Nc+1}\right]}{\Nc+1}
     -2\coupl(1-\coupl) - \frac{\coupl^2}{\Nc}
  \right]
  \\
  &   \label{eq:CovPartialWeakCoupling1}
   \underset{\coupl\ll1}{\simeq}
  -\frac{1}{\Nc^2}\times
  \left\{
  \begin{array}{ll}
    \displaystyle
    \frac{2}{\coupl \Nc} & \mbox{for } \Nc\coupl\gg1
    \\[0.25cm]
    \displaystyle
    1 & \mbox{for } \Nc\coupl\ll1
  \end{array}
  \right.
\end{eqnarray}
It is worth noting that when compared to Eq.~\eref{eq:CovPartial}, the covariances change both in sign and scaling with $\Nc$ as transmission crosses over from perfect to weak coupling. Finally,
the marginal distribution of the proper time delays at arbitrary $\coupl$ was obtained in \cite{SomSavSok01}. As will be shown in \ref{app:PartialProper}, the distributions of the proper and partial time delays become almost identical to each other in the weak coupling limit.

To close this brief overview, we emphasize an important difference between the partial (or  proper) time delays and the Wigner time delay. As is clear from the above expressions (for $\beta=2$, but the conclusion holds for any symmetry), the relative fluctuations of the partial/proper time delays are always large at weak transmission $\coupl\ll1$,
\begin{equation}
  \frac{ \mathrm{var}(\tau_a) }{ \mean{\tau_a}^2 }
  \simeq \frac{ \mathrm{var}(\tilde{\tau}_a) }{ \mean{\tilde{\tau}_a}^2 }
  \simeq \frac{2}{\coupl} \gg1.
\end{equation}
Thus one expects a broad distribution in this limit independently of the channel number $\Nc$ (see Fig.~\ref{fig:SketchPC}), as was indeed shown for the exact distributions \cite{FyoSom97,SomSavSok01} (see also~\ref{app:PartialProper} below).
On the other hand, the relative fluctuations of the Wigner time delay are not necessarily large because of the specific $N$ dependence according to \eref{eq:VarianceWT}. This can be understood from Eq.~(\ref{eq:SumRulePPW}) defining the Wigner time delay as a linear statistics on $\{\tau_a\}$ and by noting that their correlations diminish rapidly when the parameter $\Nc\coupl$ grows, see (\ref{eq:CovPartialWeakCoupling1}). Although the full distribution of the Wigner time delay is still unknown at arbitrary $T$, this discussion and (\ref{eq:VarianceWT}) suggest that it converges to the Gaussian distribution in the strong coupling limit $\Nc\coupl\gg1$. In the opposite case of weak coupling, $\Nc\coupl\ll1$, the distribution becomes broad with nontrivial behaviour. One of our purposes here is to study in much detail this broad distribution.


\section{Resonance representation for the Wigner-Smith matrix}
\label{sec:model}

\subsection{General considerations}

Our starting point is the following well-known representation for the $\Sm$-matrix in terms of an effective (non-Hermitian) Hamiltonian $\Heff$ of the open system~\cite{SokZel89,VerWeiZir85}:
\begin{equation}
  \label{eq:HamiltonianApproach}
  \Sm(\varepsilon) = \identity_\Nc-2\I\,\Wmat^\dagger \,
  \left( \varepsilon-\Heff\right)^{-1} \, \Wmat\,,
  \qquad \Heff = \mathcal{H}-\I\,\Wmat\Wmat^\dagger
  \:.
\end{equation}
This expression follows from \eref{eq:SandK} by simple algebra, but it has an advantage in making explicit the resonance energy dependence associated with the $\Sm$-matrix poles. Indeed, the latter are just given by the eigenvalue problem on $\Heff$,
$\Heff\ket{R_n}=\mathcal{E}_n\ket{R_n}$ and $\bra{L_n}\Heff=\mathcal{E}_n\bra{L_n}$, which can be further used to construct a pole expansion over the biorthogonal set of the (left and right) eigenfunctions corresponding to the same eigenvalue $\mathcal{E}_n=E_n-\frac{\I}{2}\Gamma_n$.
Since in the resonance approximation considered $\Wmat$ is assumed to be energy independent,
the energy derivative of $\Sm(\varepsilon)$ can be easily taken, leading to the following convenient representation for the Wigner-Smith matrix~\cite{SokZel97}
\begin{equation}
  \label{eq:UsefulRepresent0}
  \WSm(\varepsilon) = 2\pi \, \Psi^\dagger(\varepsilon) \, \Psi(\varepsilon)\,,
  \qquad \Psi(\varepsilon)
  =\frac{1}{\sqrt{\pi}}\, (\varepsilon-\Heff)^{-1}\, \Wmat \,.
\end{equation}
The $a$th column $\Psi_a$ of the $\Nint{\times}\Nc$ matrix $\Psi(\varepsilon)$ may be treated~\cite{SokZel97} as the internal part of the scattering wave function initiated in channel $a$ at the scattering energy $\varepsilon$. The norm of $\Psi_a$ gives the diagonal element $\WSm_{aa}$, thus providing their interpretation as the average time delay of a wave packet in a given channel~\cite{Smi60}. Using the eigenbasis of $\Heff$ and noting its completeness, we find a pole expansion of $\WSm$ as follows
\begin{equation}
  \label{eq:WS_pole}
  \WSm_{ab}(\varepsilon) = 2\sum_{n,m}
  \frac{ U_{mn} \widetilde{\Wmat}^*_{ma} \widetilde{\Wmat}_{nb} }{
  (\varepsilon-\mathcal{E}^{*}_m)(\varepsilon-\mathcal{E}_n) }\,,
\end{equation}
where $\widetilde{\Wmat}_{na}=\bra{L_n}\Wmat_{a}$ and $U_{mn}=\langle{R_m}|R_n\rangle$ is the so-called Bell-Steinberger matrix. Note that $U_{mn}\neq\delta_{mn}$ in general, so this matrix serves as a sensitive indicator of the nonorthogonality of the resonance states~\cite{FyoSav12}.

It is worth discussing
the physical meaning of the matrix $\Psi$ on an example of a quantum dot modelled by a potential.
For simplicity, we assume a discrete model and write the Hamiltonian as $\mathcal{H}_{x,x'}=-\Delta_{x,x'}+V_x\,\delta_{x,x'}$, where $\Delta$ is the discretised Laplacian matrix.
Following the same steps which have led to \eref{eq:UsefulRepresent0}, we get
\begin{equation}
   \I \left( \Sm^\dagger \derivp{\Sm}{V_x} \right)_{a,b}
   =2\pi\,\left(\Psi^\dagger\right)_{a,x}\, \Psi_{x,b}
\end{equation}
for the derivative with respect to the potential. Summation over $x$ inside the quantum dot gives \eref{eq:UsefulRepresent0}. Actually, such a formula was derived in other contexts~\cite{But00,TexBut03,TexDeg03} within a continuum model, where it was shown that
$
-(2\I\pi)^{-1} \left( \Sm^\dagger\, \delta\Sm/\delta V(x) \right)_{a,b}
=\psi^{(a)}_\varepsilon(x)^*\psi^{(b)}_\varepsilon(x)
$,
with $\psi^{(a)}_\varepsilon(x)$ being the stationary scattering state incoming in channel $a$. This leads to the correspondence
$
\Psi_{x,a}
=\frac{1}{\sqrt{\pi}}\big[(\varepsilon-\Heff)^{-1}\, \Wmat  \big]_{xa}
\equiv \psi^{(a)}_\varepsilon(x)
$
between the two models. We note, however, that taking the derivative with respect to the energy or the potential does not necessarily lead to the same result. In particular, the continuum model is known~\cite{TexBut03,TexDeg03} to have the exact relation
$
\int_\mathrm{QD}\D x\,\Sm^\dagger\, \delta\Sm/\delta V(x) =
\Sm^\dagger\, \partial_\varepsilon\Sm + \big(\Sm-\Sm^\dagger\big)/(4\varepsilon)
$,
where integration is over the scattering region (the quantum dot).
We conclude that an exact representation of $\WSm$ should not only involve $\Psi^\dagger\Psi$ like in \eref{eq:UsefulRepresent0}, but also the contribution $\big(\Sm-\Sm^\dagger\big)/(4\varepsilon)$, which is due to non-resonant effects neglected here.

Finally, it is convenient to express the Wigner-Smith matrix in terms of the reaction matrix $\Kmat$. Some algebra gives
$
\Psi=\frac{1}{\sqrt{\pi}}\,(\varepsilon-\mathcal{H})^{-1}\Wmat\,(\identity_\Nc+\I\,\Kmat)^{-1}
$,
resulting in~\cite{SokZel97}
\begin{equation}
  \label{eq:UsefulRepresent1}
  \WSm = -2\, (\identity_\Nc-\I\,\Kmat)^{-1}\,\derivp{\Kmat}{\varepsilon}\,(\identity_\Nc+\I\,\Kmat)^{-1}
  \:.
\end{equation}
This representation will prove to be useful for the RMT analysis developed below.

\subsection{RMT for perfect coupling}

The case of perfect coupling corresponds to the situation when the mean $\mean{\Sm}=0$. The $\Sm$  matrix is then distributed in one of the circular ensembles, C$\beta$E, of random orthogonal (COE, $\beta=1$), unitary ($\mathrm{CUE}\equiv\mathrm{U}(N)$, $\beta=2$) or symplectic (CSE, $\beta=4$) unitary matrices~\cite{Bee97}:
\begin{equation}
  P_\Sm^{(0)} (\Sm) \, \mathrm{D}\Sm = C_N\,\mathrm{D}\Sm\,,
\end{equation}
where $\mathrm{D}\Sm$ is the Haar measure and $C_N$ a normalisation constant (the superscript ``$^{(0)}$'' stands for perfect coupling). Correspondingly, the reaction matrix belongs to one of the three Cauchy ensembles (orthogonal, unitary or symplectic) in this case~\cite{MelPerSel85,Bro95}
\begin{equation}
  \label{eq:CauchyDistribution}
  P_\Kmat^{(0)} (\Kmat) \propto \big[\det(\identity_\Nc+\Kmat^2)\big]^{-1-\beta(\Nc-1)/2}
  \:.
\end{equation}
This follows from the relation \eref{eq:SandK} and noting that the associated Jacobian is given by
\begin{equation}
  \label{eq:JacobianSK}
  \mathrm{D}\Sm = \mathrm{D}\Kmat\,
  \frac{2^{\Nc(1+\beta(\Nc-1)/2)}}{\left[\det(\identity_\Nc+\Kmat^2)\right]^{1+\beta(\Nc-1)/2}}
  \:,
\end{equation}
where $\mathrm{D}\Kmat$ is the Lebesgue measure over the set of Hermitian matrices.

In order to derive the distribution of the Wigner-Smith matrix, we also require the statistics of the energy derivative $\partial\Kmat/\partial\varepsilon$. Following BFB \cite{BroFraBee97,BroFraBee99}, it is convenient to symmetrize the Wigner-Smith matrix according to \eref{eq:WS_sym}, which can be written as
\begin{equation}
  \WSm_{s} = -2\, (\identity_\Nc+\Kmat^2)^{-1/2}\,\derivp{\Kmat}{\varepsilon}\,(\identity_\Nc+\Kmat^2)^{-1/2}
  \:.
\end{equation}
which clearly has the same spectrum as $\WSm$. BFB's approach has shown the statistical independence of $\Kmat$ and $\partial\Kmat/\partial\varepsilon$ and, hence, that of $\Sm$ and $\WSm_{s}$, with the joint distribution
\begin{equation}
  \label{eq:BFB1997}
  P_{\Sm,\WSm_s}^{(0)}(\Sm,\WSm_{s}) = P_{\Sm}^{(0)}(\Sm)\,P_{\WSm_s}^{(0)}(\WSm_{s})
  \:.
\end{equation}
The distribution $P_{\WSm_s}^{(0)}(\WSm_s)$ turns out to correspond to a specific instance of the so-called inverse-Wishart matrices (Laguerre ensemble) for $\invQ = \WSm_s^{-1}$,
\begin{equation}
  \label{eq:Laguerre}
  P_\invQ^{(0)}(\invQ) \propto
  \Theta(\invQ)\,
  \left(\det\invQ\right)^{\beta\Nc/2} \,\EXP{-(\beta/2)\,\tr{\invQ}}
  \:.
\end{equation}
An explicit form provided by BFB for the distribution $P_{\WSm_s}^{(0)}(\WSm_s)$ of $\WSm_{s}$ then follows from the above by making use of $\mathrm{D}\invQ=(\det\WSm_s)^{-2-\beta(\Nc-1)}\,\mathrm{D}\WSm_s$.


\section{Wigner-Smith matrix distribution for non-ideal contacts}
\label{sec:WSMdistribution}

\subsection{Uniform couplings}
\label{subsec:WSMdistrib1}

We consider first a simple model of tunable contacts where all the channels are equally coupled and characterised by the same transmission coefficient $\coupl=4\fss/(1+\fss)^2$, where the coupling constant $\fss>0$ is defined in \eref{eq:RelationBetweenCouplingParameters}. The case of perfect coupling hence corresponds to $\fss=1$. In view of the resonance representation \eref{eq:HamiltonianApproach}, it is clear that the model with arbitrary coupling can be mapped to that with perfect one by performing the substitution $\Wmat\longrightarrow \sqrt{\fss}\,\Wmat$. Note that the results should depend on $\fss$ only through the transmission coefficient $T$, thus implying a symmetry $\fss\leftrightarrow1/\fss$. Such a symmetry can be understood from representation \eref{eq:SandK} and the known invariance of the Cauchy distribution \eref{eq:CauchyDistribution} under $\Kmat\leftrightarrow\Kmat^{-1}$. Therefore, it will be sufficient to consider $0<\fss\leq1$.

Keeping the notation $\Kmat$ for the reaction matrix at perfect coupling, distributed according to the Cauchy distribution \eref{eq:CauchyDistribution}, we rewrite \eref{eq:UsefulRepresent1} as follows
\begin{equation}
  \WSm =
   -2\fss\,
   (\identity_\Nc-\I\,\fss\,\Kmat)^{-1}\,
   \derivp{\Kmat}{\varepsilon}\,
   (\identity_\Nc+\I\,\fss\,\Kmat)^{-1}
   \,.
\end{equation}
Denoting the symmetrised Wigner-Smith matrix at perfect coupling by $\WSm_{s0}$, we have
\begin{equation}
  \label{eq:RelationQsQs0}
  \WSm_s= A\, \WSm_{s0}\, A\,,
  \qquad
  A = \sqrt{\fss}\,(\identity_\Nc+\fss^2\Kmat^2)^{-1/2}\,(\identity_\Nc+\Kmat^2)^{1/2}
  \:,
\end{equation}
and note also that $A=A^\dagger$. The matrix $\WSm_{s0}$ is distributed according to \eref{eq:Laguerre}. Therefore, the required distribution of $\invQ =\WSm_s^{-1} = A^{-1}  \invQ_0 A^{-1}$ can then be rewritten in terms of two integrals over Hermitian matrices from the Cauchy and Laguerre ensembles:
\begin{eqnarray}
  &\hspace{-2cm}
  P_\invQ(\invQ) = \mean{ \delta\left( \invQ - A^{-1}  \invQ_0 A^{-1} \right) }_{\Kmat,\,\invQ_0}
  \\
  \nonumber
    &\hspace{-2.5cm}
   \propto
  \int\mathrm{D}\Kmat\,\det(\identity_\Nc+\Kmat^2)^{-1-\beta(\Nc-1)/2}
  \int_{\invQ_0>0}\mathrm{D}\invQ_0\,(\det \invQ_0)^{\beta\Nc/2}\,\EXP{-(\beta/2)\tr{\invQ_0}}
  \delta\left( \invQ - A^{-1} \invQ_0 A^{-1} \right)
  \:,
\end{eqnarray}
where the second integral runs over Hermitian matrices with positive eigenvalues.
We can eliminate one matrix integral by using the general expression of the Jacobian~\cite{Mat97}
\begin{equation}
  \label{eq:UsefulJacobian}
  \mathrm{D}\invQ_0=\mathrm{D}Y\,\big[\det(A^\dagger A)\big]^{1+\beta(\Nc-1)/2}
  \hspace{1cm}\mbox{for }
  \invQ_0=A^\dagger YA
  \:,
\end{equation}
where $A$ must be real for $\beta=1$.
We finally obtain the representation
\begin{equation}
  \label{eq:DistributionZGeneral}
  \hspace{-2.5cm}
  P_\invQ(\invQ) \propto
    \Theta(\invQ)\,
    ( \det\invQ )^{\beta\Nc/2}
    \int\mathrm{D}\Kmat\,
    \frac{\det(\identity_\Nc+\Kmat^2)^{\beta\Nc/2}}
         {\det(\identity_\Nc+\fss^2\Kmat^2)^{1-\frac{\beta}{2}+\beta\Nc}}\,
    \exp\left(
      -\frac{\beta}{2}\fss\,\tr{ \frac{\identity_\Nc+\Kmat^2}{\identity_\Nc+\fss^2\Kmat^2}\invQ }
    \right)
\end{equation}
where the integration is over the set of Hermitian matrices with real ($\beta=1$), complex ($\beta=2$) or quaternionic ($\beta=4$) entries.
Setting $\fss=1$ (perfect coupling) we obviously recover the Laguerre distribution~\eref{eq:Laguerre}.

We note that in the unitary case ($\beta=2$) one can use the invariance under unitary transformations to show that the distribution of the Wigner-Smith matrix $\WSm$ is the same as the distribution of the symmetrised matrix $\WSm_s$~\cite{BroFraBee99}. However, this is not the case in the orthogonal and symplectic cases. It is tempting to perform a similar calculation as above for the $\WSm$ matrix, starting from
$\WSm = B\, \WSm_{s0}\, B^\dagger$ with
$ B = \sqrt{\fss}\,(\identity_\Nc-\I\,\fss\,\Kmat)^{-1}\,(\identity_\Nc+\Kmat^2)^{1/2} $.
This shows that the analysis done for $\WSm_s$ cannot be reproduced for $\WSm$ in the orthogonal case ($\beta=1$) because it is not clear that the change of variable $\invQ_0=B^\dagger YB$ is compatible with the constraints $\invQ_0=\invQ_0^\mathrm{T}$ and $\invQ=\invQ^\mathrm{T}$, since $\Kmat=\Kmat^\mathrm{T}$ (besides, $B$ is not real for $\beta=1$, hence \eref{eq:UsefulJacobian} cannot be used).

\subsection{Joint distribution of the eigenvalues in the unitary case ($\beta=2$)}

In the unitary case, the joint distribution of eigenvalues can be deduced from \eref{eq:DistributionZGeneral} by an integration over the unitary group. We decompose the matrices as $\invQ=V\,A\,V^\dagger$ and $\fss\frac{1+\Kmat^2}{1+\fss^2\Kmat^2}=W\,B\,W^\dagger$ where $V$ and $W$ are two unitary matrices and $A=\mathrm{diag}(\einvQ_1,\cdots,\einvQ_\Nc)$ and $B=\fss\,\mathrm{diag}(\frac{1+k_1^2}{1+\fss^2k_1^2},\cdots,\frac{1+k_\Nc^2}{1+\fss^2k_\Nc^2})$.
We have
\begin{equation}
  \mathrm{D}\invQ\,P_\invQ(\invQ) = \mathrm{D}V \, \D \einvQ_1\cdots\D \einvQ_\Nc\,
  P(\einvQ_1,\cdots,\einvQ_\Nc)\,\Delta_\Nc(\einvQ)^2
\end{equation}
where
$\Delta_\Nc(\einvQ)=\prod_{i<j}(\einvQ_i-\einvQ_j)$ is the Vandermonde and $\mathrm{D}V$ the Haar measure.
A similar decomposition holds for $\mathrm{D}\Kmat$, thus
\begin{eqnarray}
  \hspace{-2cm}
  P(\einvQ_1,\cdots,\einvQ_\Nc) \propto
  &
  \Delta_\Nc(\einvQ)^2
  \prod_n \heaviside(\einvQ_n)\,\einvQ_n^\Nc \,
  \int_{\mathbb{R}^\Nc}\D k_1\cdots\D k_\Nc\, \Delta_\Nc(k)^2
  \prod_n \frac{(1+k_n^2)^{\Nc}}{(1+\fss^2k_n^2)^{2\Nc}}
  \nonumber\\
  &\hspace{3cm}
  \times
  \int_{\mathrm{U}(\Nc)}\mathrm{D}V
  \int_{\mathrm{U}(\Nc)}\mathrm{D}W
  \EXP{-\tr{ A\,V^\dagger W\,B\,W^\dagger V } }
\end{eqnarray}
where $\heaviside(\einvQ)$ is the usual Heaviside function.
Using Harish-Chandra-Itzykson-Zuber integral (see \ref{app:HCIZ}), we obtain
\begin{eqnarray}
  \label{eq:PzArbitraryCoupling}
  \hspace{-2cm}
  P(\einvQ_1,\cdots,\einvQ_\Nc) \propto
  \Delta_\Nc(\einvQ) \,\prod_n \heaviside(\einvQ_n)\,\einvQ_n^\Nc \,
  \int_{\mathbb{R}^\Nc}\D k_1\cdots\D k_\Nc\,
  \frac{\Delta_\Nc(k)^2}{\Delta_\Nc\left(\fss\,\frac{1+k^2}{1+\fss^2k^2}\right)}
  \prod_n \frac{(1+k_n^2)^{\Nc}}{(1+\fss^2k_n^2)^{2\Nc}}
  \nonumber\\
  \hspace{6cm}
  \times
  \det\left[
    \exp\left(
      -\fss\,\frac{1+k_i^2}{1+\fss^2k_i^2}\,\einvQ_j
    \right)
  \right]
\end{eqnarray}
which will be used in Section~\ref{sec:characfct}.
\footnote{
It is worth noting that the Vandermonde determinant in the denominator can be further simplified as
\begin{equation*}
  \Delta_\Nc\left(\fss\,\frac{1+k^2}{1+\fss^2k^2}\right)
  = \left[\fss(1-\fss^2)\right]^{\Nc(\Nc-1)/2}
  \frac{\Delta_\Nc(k^2)}{\prod_n(1+\fss^2k_n^2)^{\Nc-1}}
\end{equation*}
leading to a simpler representation of equation~\eref{eq:PzArbitraryCoupling}~:
\begin{eqnarray}
  \hspace{-2.5cm}
  P(\{\einvQ_n\}) \propto
  \Delta_\Nc(\einvQ) \,\prod_n
  \heaviside(\einvQ_n)\,
  \einvQ_n^\Nc \,
  \int
  \D k_1\cdots\D k_\Nc\,
  \frac{\Delta_\Nc(k)^2}{\Delta_\Nc (k^2)}
  \prod_n \frac{(1+k_n^2)^{\Nc}}{(1+\fss^2k_n^2)^{\Nc+1}}
  \nonumber
  \det\left[
    \exp\left(
      -\fss\,\frac{1+k_i^2}{1+\fss^2k_i^2}\,\einvQ_j
    \right)
  \right].
\end{eqnarray}
}

\subsection{Channels with different coupling parameters }
\label{subsec:DifferentCouplings}

It is clear from the above discussion how to extend the obtained results to a general case of arbitrary and nonequal channel couplings. Exploiting representation \eref{eq:SandK} again, we can now substitute the reaction matrix $\Kmat$ at perfect couplings by
$$
  \Kmat \longrightarrow  \mathcal{U}_C^\dagger \, C \, \Kmat \, C\, \mathcal{U}_C
  \,, \qquad
  C=\mathrm{diag}(\sqrt{\fss_1},\cdots,\sqrt{\fss_\Nc})\,,
$$
where $\mathcal{U}_C$ is a unitary matrix and $\fss_a$ correspond to different transmission coefficients \eref{eq:RelationBetweenCouplingParameters}.
Following the same lines as in section~\ref{subsec:WSMdistrib1}, we have $\WSm_s=A\,\WSm_{s0}\,A^\dagger$ with
\begin{equation}
  A = \mathcal{U}_C^\dagger\,
  \left( \identity_\Nc + \left(C\,\Kmat\,C\right)^2 \right)^{-1/2}\, C\,
  \left( \identity_\Nc+\Kmat^2 \right)^{1/2}
  \:.
\end{equation}
(We have used $\big[\mathcal{U}_C^\dagger M\mathcal{U}_C\big]^{-1/2}=\mathcal{U}_C^\dagger M^{-1/2}\mathcal{U}_C$, but note that $(AB)^{-1/2}\neq B^{-1/2}A^{-1/2}$ in general).
As before, we assume that all energy dependence is carried by the reaction matrix $\Kmat$, while the matrices $\mathcal{U}_C$ and $C$ of the coupling parameters are energy independent.
This leads to the following generalisation of equation~\eref{eq:DistributionZGeneral}:
\begin{eqnarray}
  \label{eq:DistribGammaArbitraryCouplings}
\hspace{-2cm}
  P_\invQ(\invQ) \propto
    \Theta(\invQ)\,
    ( \det\invQ )^{\beta\Nc/2}
    \int\mathrm{D}\Kmat\,
    \frac{\det(\identity_\Nc+\Kmat^2)^{\beta\Nc/2}}
         {\det\left(\identity_\Nc+\left(C\Kmat C\right)^2\right)^{1-\frac{\beta}{2}+\beta\Nc}}
         \\
\hspace{-2cm}
    \nonumber
    \times
    \exp\left(
      -\frac{\beta}{2}
      \tr{
         \left( \identity_\Nc + \left(C\,\Kmat\,C\right)^2  \right)^{-1/2}\,C\,
         \left( \identity_\Nc + \Kmat^2\right)\, C\,
         \left( \identity_\Nc + \left(C\,\Kmat\,C\right)^2  \right)^{-1/2}\,
         \mathcal{U}_C\,\invQ\,\mathcal{U}_C^\dagger
         }
    \right).
\end{eqnarray}
One obviously recovers \eref{eq:DistributionZGeneral} at $\mathcal{U}_C=\identity_\Nc$ and $C=\sqrt{\fss}\,\identity_\Nc$.


\section{Joint distribution of $\Sm$ and $\WSm_s$ (uniform couplings)}
\label{sec:IntegrationOverCUE}

\begin{figure}[!ht]
\centering
\includegraphics[height=3.5cm]{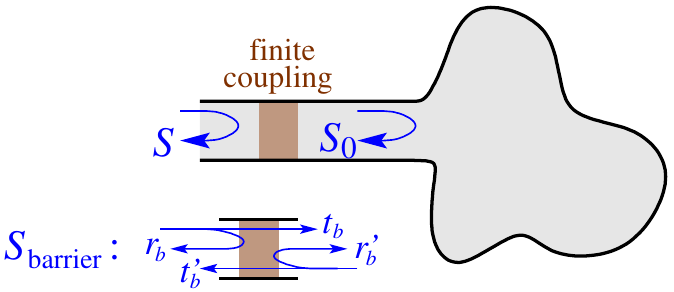}
\hspace{1cm}
\includegraphics[height=3.5cm]{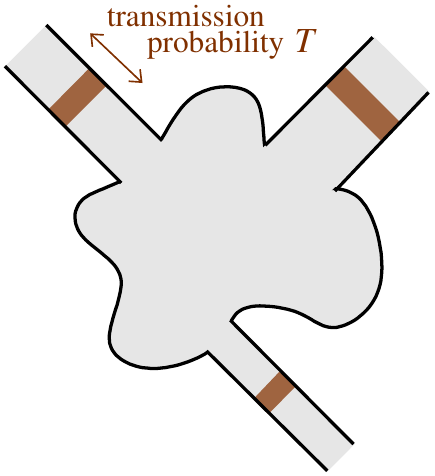}
\caption{\it Quantum dots coupled to contacts through which the electronic wave is injected.
  The scattering matrix $\Sm_0$ describes the dynamics of the perfectly coupled quantum dots and the scattering matric $\Sm_\mathrm{barrier}$ the scattering through each barrier.
}
\label{fig:QDcouplings}
\end{figure}

We now derive another instructive integral representation for the distribution $P_\invQ(\invQ)$ in terms of an integral over the unitary group.
Our purpose here is not simply technical but aiming to shed new light on the derivation of Eq.~\eref{eq:DistributionZGeneral}. This second formulation will allow us to obtain more straightforwardly the joint distribution $P(\Sm,\invQ)$ of the matrices $\Sm$ and $\invQ$. It will also be useful for the numerical calculations presented in Section~\ref{sec:Numerics}. The starting point is to reformulate the model introduced above, according to Brouwer's construction~\cite{Bro95} of the distribution of the $\Sm$ matrix for tunable couplings. We introduce the $\Nc\times\Nc$ scattering matrix $\Sm_0$ belonging to one of the circular ensembles C$\beta$E, describing the quantum dot for perfect contacts.
The non-ideal nature of the contact is then accounted for through the $2\Nc\times2\Nc$ scattering matrix (see Fig.~\ref{fig:QDcouplings})
\begin{equation}
  \Sm_\mathrm{barrier} =
   \left(
  \begin{array}{cc}
    r_b & t_b' \\ t_b & r_b'
  \end{array}
  \right)
\end{equation}
gathering the transmission/reflection through the region between the lead and the dot.

\subsection{The scattering matrix}

The matrix ${\Sm_0}$ is such that $\mean{\Sm_0}=0$ by construction, while the matrix $\Sm_\mathrm{barrier}$ is supposed fixed. The scattering matrix describing the quantum dot with arbitrary couplings is
\begin{equation}
  \Sm = r_b + t_b' \, \left( \Sm_0^\dagger - r_b' \right)^{-1} \, t_b
  \:.
\end{equation}
Because $\mean{\Sm_0^n}=0$ for any positive integer $n$, we have $\mean{\Sm}=r_b$. We still consider the case of uniform couplings, when the average $\mean{\Sm}\equiv\mathbf{1}_\Nc\,\Sbar$ is proportional to the identity matrix,
\begin{equation}
  r_b=-\left(r_b'\right)^\dagger=\mathbf{1}_\Nc\,\Sbar
  \hspace{1cm}\mbox{and}\hspace{1cm}
  t_b=t_b'=\mathbf{1}_\Nc\,\sqrt{1-|\Sbar|^2}
  \:,
\end{equation}
leading to the simpler representation
\begin{equation}
  \label{eq:TheModel}
  \Sm =
  \Big( \Sbar\,\mathbf{1}_\Nc + \Sm_0 \Big)\, \Big(  \mathbf{1}_\Nc + \Sbar^*\,\Sm_0 \Big)^{-1}
  \:.
\end{equation}
Introducing the transmission probability $\coupl=1-|\Sbar|^2$ of the barrier, we note that the case of perfect couplings, $\coupl=1$, corresponds to $\Sm=\Sm_0$ with $\mean{\Sm}\equiv\Sbar=0$.

Given these results, we can now obtain the distribution of $\Sm$ at arbitrary coupling by evaluating the Jacobian of transformation~\eref{eq:TheModel}. Note that the two scattering matrices have the same eigenvectors. Then establishing a relation between the two measures only requires to relate the Vandermonde determinants constructed from their eigenvalues.
Using
\begin{equation}
  \Sm_0=\Big( \Sm - \Sbar\,\mathbf{1}_\Nc \Big)\,\Big( \mathbf{1}_\Nc-\Sbar^*\,\Sm \Big)^{-1}
  \:,
\end{equation}
we deduce the following relation between eigenvalues
\begin{equation}
      \EXP{\I\theta_a^{(0)}} - \EXP{\I\theta_b^{(0)}}
    = \frac{1-|\Sbar|^2}{(1-\Sbar^*\,\EXP{\I\theta_a})(1-\Sbar^*\,\EXP{\I\theta_b})}
    \big[ \EXP{\I\theta_b} - \EXP{\I\theta_a} \big]
    \:.
  \end{equation}
As a consequence, the Vandermondes built from the two sets of eigenvalues are related by
\begin{equation}
\hspace{-2cm}
      \Delta_\Nc\big( \EXP{\I\theta_1^{(0)}},\cdots,\EXP{\I\theta_\Nc^{(0)}} \big)
    = \big(1-|\Sbar|^2\big)^{\Nc(\Nc-1)/2}
    \prod_a\big(1-\Sbar\,\EXP{\I\theta_a}\big)^{-\Nc+1}\,
    \Delta_\Nc\big( \EXP{\I\theta_1},\cdots,\EXP{\I\theta_\Nc} \big)\,.
  \end{equation}
Using $\D\theta_a^{(0)}=\D\theta_a\,(1-|\Sbar|^2)/|1-\Sbar^*\,\EXP{\I\theta_a}|^2$, we finally relate the two measures as follows
\begin{equation}
  \mathrm{D}\Sm_0 =
  (1-|\Sbar|^2)^{\Nc+\beta\Nc(\Nc-1)/2}\,
  \mathrm{D}\Sm
  \,|\det(\identity_\Nc-\Sbar^*\Sm)|^{-2-\beta(\Nc-1)}
\end{equation}
from which we can read out the distribution $P_{\Sm}(\Sm)$ of $\Sm$, as $P_{\Sm}^{(0)}(\Sm)$ is just uniform. We have thus recovered the Poisson kernel (\ref{eq:PoissonKernel}) (reproducing the proof of Ref.~\cite{Bro95}).

We now consider the Wigner-Smith time matrix.
Assuming as before $\partial_\varepsilon\Sbar=0$, we have
$
   \partial_\varepsilon\Sm
   =\left(1-|\Sbar|^2\right)\,
   \big( \identity_\Nc + \Sbar^*\Sm_0 \big)^{-1} \,
   \partial_\varepsilon\Sm_0\,
   \big( \identity_\Nc + \Sbar^*\Sm_0 \big)^{-1}
$,
thus yielding
\begin{equation}
  \WSm = \left( 1-|\Sbar|^2 \right) \,
  \left( \identity_\Nc + \Sbar\Sm_0^\dagger  \right)^{-1} \,
  \WSm_0\,
  \left( \identity_\Nc + \Sbar^*\Sm_0  \right)^{-1}
  \:.
\end{equation}
The symmetrised Wigner-Smith matrix can be again written as $\WSm_s = A\,\WSm_{s0}\,A$, where
\begin{equation}
  \label{eq:QA}
  A = \sqrt{1-|\Sbar|^2}
    \left( \identity_\Nc + \Sbar^*\Sm_0  \right) ^{-1/2}
    \left( \identity_\Nc + \Sbar\Sm_0^\dagger  \right) ^{-1/2}
\end{equation}
is Hermitian. One can easily check that this expression is equivalent to \eref{eq:RelationQsQs0}.

It will be useful in what follows to also determine the Jacobian of the transformation $\invQ_0\to\invQ=A^{-1}\invQ_0(A^\dagger)^{-1}$.
Using \eref{eq:UsefulJacobian} we obtain
\begin{equation}
  \mathrm{D}\invQ_0 =
    (1-|\Sbar|^2)^{\Nc+\beta\Nc(\Nc-1)/2}
  \,\left|\det\left(\identity_\Nc+\Sbar^*\Sm_0)\right)\right|^{-2-\beta(\Nc-1)}\,
  \mathrm{D}\invQ
\end{equation}
which can be re-expressed in terms of $\Sm$, leading to
\begin{equation}
  \mathrm{D}\invQ_0 =
  (1-|\Sbar|^2)^{-\Nc-\beta\Nc(\Nc-1)/2}\,
  \left|\det\left(\identity_\Nc-\Sbar^*\Sm\right)\right|^{2+\beta(\Nc-1)}\,
  \mathrm{D}\invQ
  \:.
\end{equation}
Remarkably, this shows that the measure is invariant,
\begin{equation}
  \label{eq:ConservationMeasureSinvQ}
  \mathrm{D}\Sm_0\, \mathrm{D}\invQ_0  =  \mathrm{D}\Sm\, \mathrm{D}\invQ
\end{equation}
It is tempting to regard this equation as a matrix extension of Liouville's theorem, although further study would be needed to support this statement (e.g., by investigating parametric evolution of the associated matrix flow with regard to coupling changes).

\subsection{Joint distribution of $\Sm$ and $\invQ=\WSm_s^{-1}$}

Our starting point is again the BFB result for ideal contacts \eref{eq:BFB1997}, rewritten with the inverse Wigner-Smith matrix
\begin{equation}
  P^{(0)}_{\Sm,\invQ}(S_0,\invQ_0)
  \propto
  \Theta(\invQ_0)\,
  \left(\det\invQ_0\right)^{\beta\Nc/2} \,\EXP{-(\beta/2)\,\tr{\invQ_0}}
\end{equation}
Using the two transformations (\ref{eq:TheModel}) and (\ref{eq:QA}), and the conservation of the measure \eref{eq:ConservationMeasureSinvQ}, we finally obtain the joint distribution
\begin{eqnarray}
  \label{eq:JointPDFSinvQ}
  &&
  \hspace{-1.75cm}
  P_{\Sm,\invQ}(\Sm,\invQ)
  \propto
  \Theta(\invQ)\,
  \left|\det\big(\identity_\Nc-\Sbar^*\Sm\big)\right|^{\beta\Nc}\,
  \\\nonumber
  &&\times
  \big(\det\invQ\big)^{\beta\Nc/2} \,
  \exp\left[
     -\frac{\beta}{2(1-|\Sbar|^2)}\,
     \tr{(\identity_\Nc-\Sbar^*\Sm)(\identity_\Nc-\Sbar\Sm^\dagger)\invQ}
  \right]
\end{eqnarray}
A similar structure was given in a recent paper~\cite{MarSchBee16}, including the other four (``BdG'') symmetry classes relevant for scattering in an Andreev billiard.
The case of the three chiral symmetry classes remains an open problem.

\subsection{Distribution of the inverse of the Wigner-Smith matrix}

The distribution of the matrix $\invQ = \WSm_s^{-1}$ can be deduced by integrating \eref{eq:JointPDFSinvQ} over $\Sm$.
In order to make the connection with the representation \eref{eq:DistributionZGeneral} more clear, we prefer to write a matrix integral over the scattering matrix of the cavity with perfect contacts:
\begin{equation}
  P_\invQ(\invQ) =
  \mean{ \delta\left( \invQ - (A^{-1})^\dagger \invQ_0 A^{-1} \right) }_{\Sm_0,\,\invQ_0}
  \:,
\end{equation}
where $\Sm_0$ belongs to the circular ensemble and $\invQ_0=\WSm_{s0}^{-1}$ is uncorrelated from the scattering matrix and distributed according to \eref{eq:Laguerre}.
Using \eref{eq:UsefulJacobian}, we finally obtain
\begin{eqnarray}
  \label{eq:DistribWSoverUnitaryGroup}
\hspace{-1cm}
  P_\invQ(\invQ) \propto
   \Theta(\invQ)\,
  &( \det\invQ )^{\beta\Nc/2}
  \int_{\mathrm{C\beta E}}
  \mathrm{D}\Sm_0\,
  \left|\det\left(\identity_\Nc+\Sbar^*\Sm_0\right)\right|^{\beta-2-2\beta\Nc}
  \\
  \nonumber
  &\times
  \exp\left[
    -\frac{\beta}{2}\,(1-|\Sbar|^2)
    \tr{( \identity_\Nc + \Sbar^*\Sm_0)^{-1}( \identity_\Nc + \Sbar\Sm_0^\dagger)^{-1} \invQ }
    \right]
    \:,
\end{eqnarray}
where the integral runs over the circular ensemble.
Note that it is also possible to go more directly from \eref{eq:DistributionZGeneral} to \eref{eq:DistribWSoverUnitaryGroup} by using \eref{eq:SandK} and \eref{eq:JacobianSK}.

The generalisation of this result to the case of channels with different couplings, as it has been done in Section~\ref{subsec:DifferentCouplings}, is also possible.


\section{Characteristic function of the Wigner time delay}
\label{sec:characfct}

As is already mentioned in the introduction, the trace of the Wigner-Smith matrix
\begin{equation}
  \Wt = \frac{1}{\Nc}\sum_a\tau_a  =\frac{1}{\Nc}\tr{ \invQ^{-1} }\,,
\end{equation}
i.e. the Wigner time delay, is of special interest due to its practical applications. The distribution and moments of $\Wt$ were studied in much detail for perfect coupling $\coupl=1$ \cite{GopMelBut96,FyoSom97,SavFyoSom01,MezSim13,TexMaj13}.
Our aim now is to determine the distribution $\dwt{\Nc}{\beta}(\tau)$ of the Wigner time delay in the weak coupling limit $\coupl\approx 4\fss\to0$. We find it convenient to introduce the rescaled variable $t = 2\Wt/(\beta\fss)$, with the rescaled distribution being
\begin{equation}
  \label{eq:DefRescaledDistWTD}
  \Rdwt{\Nc}{\beta}(t)=(\beta\fss/2)\,\dwt{\Nc}{\beta}(\tau=(\beta\fss/2)\,t)
\end{equation}
(we will see in Section~\ref{sec:Numerics} and \ref{app:PartialProper} that a more natural scaling variable is $|1/\fss-\fss|\,\tau$ rather than $\tau/\fss$, however this makes no difference in the weak coupling limit). We introduce the characteristic function for the Wigner time delay
\begin{equation}
  \label{eq:DefZN}
  \mathcal{Z}_{\Nc,\beta}(p)
  =
  \mathcal{Z}_{\Nc,\beta}(0)\,  \mean{ \EXP{ -(2p/\beta\fss)\tr{\invQ^{-1}} }  }
\end{equation}
(the normalisation $\mathcal{Z}_{\Nc,\beta}(0)$ will be chosen for convenience below). The characteristic function is related to the distribution of the rescaled time delay as
\begin{equation}
  \frac{\mathcal{Z}_{\Nc,\beta}(p)}{\mathcal{Z}_{\Nc,\beta}(0)}
  = \int_0^\infty\D\rt\,
  \Rdwt{\Nc}{\beta}(t)\, \EXP{-\Nc p\rt}
  \:.
\end{equation}

In the following, we mostly consider the unitary case $\beta=2$.
The last subsection will discuss the large deviation for arbitrary symmetry class.
The characteristic function can be written as a matrix integral with \eref{eq:DistributionZGeneral}. Using expression \eref{eq:PzArbitraryCoupling} for the joint distribution of the eigenvalues $\{\einvQ_1,\cdots,\einvQ_\Nc\}$, we get
\begin{eqnarray}
  \hspace{-2cm}
  \mathcal{Z}_{\Nc,2}(p)
  \propto
  \int_{\mathbb{R}_+^\Nc}\D \einvQ_1\cdots\D \einvQ_\Nc\,
  &\Delta_\Nc(\einvQ) \,
  \prod_n \einvQ_n^\Nc \,\EXP{-p/(\fss \einvQ_n)}
  \int_{\mathbb{R}^\Nc}\D k_1\cdots\D k_\Nc\,
  \frac{\Delta_\Nc(k)^2}{\Delta_\Nc\left(\fss\,\frac{1+k^2}{1+\fss^2k^2}\right)}
  \nonumber\\
  &\times
  \prod_n \frac{(1+k_n^2)^{\Nc}}{(1+\fss^2k_n^2)^{2\Nc}}
  \det\left[
    \exp\left(
      -\fss\,\frac{1+k_i^2}{1+\fss^2k_i^2}\,\einvQ_j
    \right)
  \right]
\end{eqnarray}
Integrals over $\gamma_k$ can be performed thanks to the Andreief formula (see \ref{app:Andreief})
\begin{eqnarray}
  \hspace{-1cm}
  \int_{\mathbb{R}_+^\Nc}
  \prod_k\left(\D \einvQ_k\, \einvQ_k^\Nc \,\EXP{-p/(\fss \einvQ_k)} \right)
  \det\left[ \einvQ_k^{i-1} \right]
  \,
  \det\left[
    \exp\left(
      -\fss\,\einvQ_k\,\frac{1+k_j^2}{1+\fss^2k_j^2}
    \right)
  \right]
  \nonumber\\
  =\Nc!\,
  \det\left[
    \int_0^\infty\D \einvQ\, \einvQ^\Nc \,\EXP{-p/(\fss \einvQ)}
    \einvQ^{i-1} \,
    \EXP{-\fss\,\einvQ\,(1+k_j^2)/(1+\fss^2k_j^2)}
  \right]
\end{eqnarray}
leading to
\begin{eqnarray}
  \hspace{-2cm}
  \mathcal{Z}_{\Nc,2}(p)
  =\int_{\mathbb{R}^\Nc}\D k_1\cdots\D k_\Nc\,
  \frac{\Delta_\Nc(k)^2}{\Delta_\Nc(\frac{1+k^2}{1+\fss^2k^2})}\,
  \prod_n
  \frac{(1+k_n^2)^{\Nc}}{(1+\fss^2k_n^2)^{2\Nc}}\,
  \nonumber\\
  \hspace{2cm}\times
  \det\left[
    \left(
       p\,
       \frac{1+\fss^2k_j^2}{1+k_j^2}
    \right)^{\frac{\Nc+i}{2}}
    K_{\Nc+i} \left(2\sqrt{p\frac{1+k_j^2}{1+\fss^2k_j^2}}\right)
  \right]
  \:.
\end{eqnarray}
This expression can be simplified further by noticing the obvious relation
\begin{equation}
\label{eq:UsefulRelation}
  \prod_j\xi_j^{2\Nc}\,\det\left[ \xi_j^{-\Nc-i} \right] = \Delta_\Nc(\xi)
  \:, \qquad
  \xi_j \equiv \frac{1+k_j^2}{1+\fss^2k_j^2}
  \,.
\end{equation}
Collecting everything, we arrive at the final result
\begin{eqnarray}
  \label{eq:MainResult1}
  \hspace{-2.5cm}
  \mathcal{Z}_{\Nc,2}(p)
  =\int_{\mathbb{R}^\Nc}\D k_1\cdots\D k_\Nc\,
  \frac{ \Delta_\Nc(k)^2 }{ \prod_n (1+k_n^2)^{\Nc} }
  \,
  \frac{
  \det\left[
    \left(
       p\,
       \frac{1+\fss^2k_j^2}{1+k_j^2}
    \right)^{\frac{\Nc+i}{2}}
    K_{\Nc+i} \left(2\sqrt{p\frac{1+k_j^2}{1+\fss^2k_j^2}}\right)
  \right]
  }{
  \det\left[
    \left(
       \frac{1+k_j^2}{1+\fss^2k_j^2}
    \right)^{-\Nc-i}
  \right]
  }
\end{eqnarray}

\paragraph{Normalisation constant.}

Using the asymptotics of the MacDonald function,
$K_\nu(x)\simeq\big[\Gamma(\nu)/2\big]\,(2/x)^\nu$ for $x\to0$,
we get the normalisation constant under the form
\begin{equation}
  \mathcal{Z}_{\Nc,2}(0)
  =2^{-\Nc}\prod_{n=1}^\Nc\Gamma(\Nc+n)
  \int_{\mathbb{R}^\Nc}\D k_1\cdots\D k_\Nc\,
  \Delta_\Nc(k)^2\prod_n(1+k_n^2)^{-\Nc}
  \:,
\end{equation}
which is surprisingly independent of $\fss$. We recognize the normalisation of the Cauchy ensemble, Eq.~\eref{eq:mi1} of \ref{app:tmi}, hence we get
\begin{equation}
  \mathcal{Z}_{\Nc,2}(0)
  =\pi^\Nc 2^{-\Nc^2}\Nc! \prod_{n=1}^\Nc\Gamma(\Nc+n)
  \:.
\end{equation}

\subsection{Perfect coupling}

Eq.~\eref{eq:MainResult1} shows that the limit of perfect coupling, $\fss\to1$, is singular as the determinant in the denominator vanishes. For this reason it is more easy to start from the definition \eref{eq:DefZN} with \eref{eq:Laguerre} and apply the Andreief formula with
\begin{equation}
  \mathcal{Z}_{\Nc,2}(p) \propto
  \int_{\mathbb{R}_+^\Nc}\D\einvQ_1\cdots\D\einvQ_\Nc\,
  \Delta_\Nc(\einvQ)^2\,\prod_k\left( \einvQ_k^\Nc \, \EXP{-\einvQ_k-p/\einvQ_k} \right)
\end{equation}
leading to
\begin{equation}
  \mathcal{Z}_{\Nc,2}(p)
  \propto
  \det\left[
    p ^{\frac{\Nc+i+j-1}{2}}
    \,
    K_{\Nc+i+j-1}(2\sqrt{p})
  \right]
  \hspace{1cm}\mbox{for }
  \fss=1
  \:.
\end{equation}
For two other symmetry classes ($\beta=1$, $4$), one can also obtain certain Pfaffian representation (analogous to the one derived in a different context in Ref.~\cite{GraTex16}, cf. supplementary material to this paper as well as \cite{Gra18}).

\subsection{Limiting behaviours of the characteristic function in the weak coupling limit}
\label{subsec:UniversalLimitWC}

The form \eref{eq:MainResult1} is appropriate to consider the weak coupling limit $\fss\to0$~:
the characteristic function \eref{eq:MainResult1} simplifies as
\begin{equation}
  \label{eq:CharacFctWC}
  \hspace{-1cm}
  \mathcal{Z}_{\Nc,2}(p)
  =\int_{\mathbb{R}^\Nc}\D k_1\cdots\D k_\Nc\,
  \Delta_\Nc(k)^2 \:
  \frac{
  \det\left[
    \left(
       \frac{p}{1+k_j^2}
    \right)^{\frac{\Nc+i}{2}}
    K_{\Nc+i} \left(2\sqrt{p(1+k_j^2)}\right)
  \right]
  }{
  \det\left[
     (1+k_j^2)^{-i}
  \right]
  }
  \:.
\end{equation}
The existence of a finite limit for $\fss\to0$ shows that the distribution $\dwt{\Nc}{\beta}(\tau)$ admits a universal form (independent of the coupling) after proper rescaling $\tau\sim\fss$, i.e. the rescaled distribution $\Rdwt{\Nc}{\beta}(t)$ has a limit.
A similar observation is made for the marginal distributions of both partial and proper time delays in~\ref{sec:Partial} for arbitrary symmetry class.

\subsubsection{Limit $p\to\infty$.}

In the limit  $p\to\infty$, the determinant \eref{eq:MainResult1} simplifies to
\begin{equation}
  \label{eq:PinftyStep1}
  \hspace{-2cm}
  \mathcal{Z}_{\Nc,2}(p)
  \simeq
  \left(\frac{\pi}{4}\right)^{\Nc/2}
  p^{\frac{3\Nc^2}{4}}
  \int\D k_1\cdots\D k_\Nc\, \Delta_\Nc(k)^2 \,
  \prod_n\left[
    \frac{ \EXP{-2\sqrt{p(1+k_n^2)}} }{(1+k_n^2)^{\frac\Nc2+\frac14}}
    \right]
    \frac{
      \det\left[ (1+k_j^2)^{-i/2}  \right]
         }{
      \det\left[ (1+k_j^2)^{-i}  \right]
         }
  \:.
\end{equation}
The exponentials constraint the variables to be $k_n\lesssim1/\sqrt{p}\to0$, thus we can write
$\EXP{-2\sqrt{p(1+k_n^2)}}\simeq\EXP{-2\sqrt{p}-\sqrt{p}k_n^2}$ and expand the remaining functions.

We now analyse the ratio of the two determinants in the limit $k_j\to0$. For this purpose, we use the following convenient relation
\begin{equation}
  \label{eq:UsefulRelationWithDet}
  \det\left[
    \phi_i(k_j)
  \right]_{1\leq i,\, j\leq N}
  \underset{k_j\to0}{\simeq} \Delta_N(k) \,
  \det\left[
    \phi_i^{(n-1)}(0)/(n-1)!
  \right]_{1\leq i,\, n\leq N}
\end{equation}
where $\{\phi_i(k)\}$ is a set of regular functions (differentiable at least $N$ times).
The proof of the relation is simple: replacing the functions by a Taylor expansion, we notice that the lowest order in $k_j$'s is provided by the first $N$ terms of the series
$$
  \det\left[
    \sum_{n=1}^{N-1}\frac{\phi_i^{(n-1)}(0)}{(n-1)!} \, k_j^{n-1}
  \right]_{1\leq i,\, j\leq N}\,.
$$
This is readily recognized as the determinant of a product of matrices, yielding \eref{eq:UsefulRelationWithDet}.

We apply \eref{eq:UsefulRelationWithDet} to the ratio of determinants in \eref{eq:PinftyStep1}.
The corresponding Taylor expansion is given by
$(1+x)^{-\alpha}=\sum_{n=0}^\infty\frac{(\alpha)_n}{n!}(-x)^n$, where $(\alpha)_n=\Gamma(\alpha+n)/\Gamma(\alpha)$ is the Pochhammer symbol.
Thus
the ratio of the two determinants has a finite limit
\begin{equation}
      \frac{
      \det\left[ (1+k_j^2)^{-i/2}  \right]
         }{
      \det\left[ (1+k_j^2)^{-i}  \right]
         }
   \underset{ k_j\to0 }{\longrightarrow}
   \mathscr{B}_\Nc =
  \frac{ \det\left[ (-1)^{j-1}\,\frac{\Gamma(i/2+j-1)}{\Gamma(i/2)\Gamma(j)} \right] }
       { \det\left[ (-1)^{j-1}\,\frac{\Gamma(i+j-1)}{\Gamma(i)\Gamma(j)} \right] }\,,
\end{equation}
which after further simplifications reduces to
\begin{equation}
  \mathscr{B}_\Nc =
  \frac{ \det\left[ \Gamma(i/2+j-1) \right] }{ \det\left[ \Gamma(i+j-1) \right] }
  \,\prod_{n=1}^\Nc\frac{\Gamma(n)}{\Gamma(n/2)}\,.
\end{equation}

We can now write
\begin{equation}
  \mathcal{Z}_{\Nc,2}(p)
  \simeq
  \left(\frac{\pi}{4}\right)^{\Nc/2}\mathscr{B}_\Nc\,
  p^{\frac{3\Nc^2}{4}} \EXP{-2\Nc\sqrt{p}}
  \int\D k_1\cdots\D k_\Nc\,\Delta_\Nc(k)^2
  \prod_n \EXP{-\sqrt{p}k_n^2}\,.
\end{equation}
Using $\Delta_N(\alpha\,x)=\alpha^{N(N-1)/2}\,\Delta_N(x)$ and the integral \eref{eq:mi2}, we finally obtain
\begin{equation}
  \label{eq:CaraFctWTDBeta2}
  \mathcal{Z}_{\Nc,2}(p)
  \underset{p\to\infty}{\simeq}
  \mathscr{A}_\Nc\,
  p^{\frac{\Nc^2}{2}} \, \EXP{-2\Nc\sqrt{p}}\,,
\end{equation}
where $\mathscr{A}_\Nc = 2^{-\frac{\Nc^2}{2}}\left(\frac{\pi}{4}\right)^{\Nc/2} \mathscr{B}_\Nc \int\D x_1\cdots\D x_\Nc\,\Delta_\Nc(x)^2\prod_n \EXP{-x_n^2/2}$
can be also written as
\begin{eqnarray}
  \mathscr{A}_\Nc
   = 2^{-\frac{\Nc}{2}(\Nc+1)}\pi^{\Nc}\,G(N+2)\,\mathscr{B}_\Nc
\end{eqnarray}
in terms of the Barnes $G$-function.

Correspondingly, the (rescaled) Wigner time delay distribution reads
\begin{equation}
    \Rdwt{\Nc}{2}(\rt)
    \underset{t\to0}{\simeq}  \mathscr{C}_\Nc\,
    \rt^{-\Nc^2-3/2}\,\EXP{-\Nc/t}\,,
    \qquad
    \mathscr{C}_\Nc =\sqrt{\frac{\Nc}{\pi}} \frac{\mathscr{A}_\Nc}{\mathcal{Z}_{\Nc,2}(0)}
    \:,
\end{equation}
thus yielding the asymptotic behaviour
\begin{equation}
  \dwt{\Nc}{2}(\tau)  \sim
  \coupl^{-1}\,(\coupl/\tau)^{\Nc^2+3/2}
  \,\EXP{-\Nc\coupl/(4\tau)} \qquad\mbox{for } \tau\ll\coupl
  \:.
\end{equation}

\subsubsection{Limit $p\to0$.}

The limit of small $p$ is more tricky.
First, it must be recognised that the dominant contribution to the multiple integral \eref{eq:CharacFctWC} comes from the expansion of the MacDonald functions within a window $|k_n|\lesssim1/\sqrt{p}$~:
$$
    \mathcal{Z}_{\Nc,2}(p)
  =
  \frac{\prod_n\Gamma(\Nc+n)}{2^{\Nc}}
  \int\D k_1\cdots\D k_\Nc\,
  \frac{ \Delta_\Nc(k)^2 }{ \prod_n (1+k_n^2)^{\Nc} }
  \,
  \frac{
   \det\left[
       \frac{1}{(1+k_j^2)^i}
       \left(
          1 - p\, \frac{1+k_j^2}{\Nc+i-1} + \mathcal{O}(p^2)
       \right)
  \right]
    }{\det\left[ (1+k_j^2)^{-i}  \right]}
$$
Now we use that the $p$-dependent determinant here can be further written as
\begin{equation}
  \det(A - p\,B) \underset{p\to0}{\simeq} \det(A) \left( 1 - p\, \tr{A^{-1}B} \right)\,,
\end{equation}
where the matrices $A$ and $B$ are defined by
\begin{equation}
  A_{ij}=(1+k_j^2)^{-i}\equiv(X_j)^{i}
  \quad\mbox{and}\quad
  B_{ij}=\frac{(1+k_j^2)^{-i+1}}{\Nc+i-1}\equiv\frac{(X_j)^{i-1}}{\Nc+i-1}
  \:.
\end{equation}
Making use of the relation
\begin{equation}
    \tr{A^{-1}B}  = \frac{1}{\Nc} \sum_n X_n^{-1} = \frac{1}{\Nc} \sum_n (1+k_n^2)
    \:,
\end{equation}
the leading order term of the characteristic function is then found as follows
\begin{eqnarray}
 \hspace{-2cm}
   \mathcal{Z}_{\Nc,2}(0) -  \mathcal{Z}_{\Nc,2}(p)
   \sim p   \int_{|k_i|\lesssim1/\sqrt{p}}\D k_1\cdots\D k_\Nc\,
   \Delta_\Nc(k)^2
   \prod_{n}(1+k_n^2)^{-\Nc}
   \frac{1}{\Nc}\sum_n (1+k_n^2)
\end{eqnarray}
where we have used once again
$\det\big[(1+k_j^2)^{-i}\big]=\Delta_\Nc(k^2)\, \prod_{n}(1+k_n^2)^{-\Nc}$.
By symmetry we can perform $(1/\Nc)\sum_n(1+k_n^2)\to 1+ k_\Nc^2$.
As $p\to0$, the dominant contribution comes from the term
\begin{equation}
   \Delta_\Nc(k)^2
  \simeq k_\Nc^{2(\Nc-1)}  \Delta_{\Nc-1}(k)^2 + \mathcal{O}( k_\Nc^{2\Nc-3} )\,.
\end{equation}
By inspecting the integral, we can write
$$
   \mathcal{Z}_{\Nc,2}(0) -  \mathcal{Z}_{\Nc,2}(p)
   \sim p
   \underbrace{
   \int_{|k_i|\lesssim1/\sqrt{p}}\D k_1
   \cdots\D k_{\Nc-1}\,
  \frac{ \Delta_{\Nc-1}(k)^2}{ \prod_{n=1}^{\Nc-1}(1+k_n^2)^{\Nc} }
   }_{\to \mathrm{const.} \ \mathrm{as}\  p\to0}
   \underbrace{
    \int_{|k_\Nc|\lesssim1/\sqrt{p}}\D k_\Nc \, \frac{ k_\Nc^{2(\Nc-1)}\,(1+k_\Nc^2)}{(1+k_\Nc^2)^\Nc}
   }_{\sim 1/\sqrt{p} \ \mathrm{as}\  p\to0}
$$
and, therefore, conclude that
\begin{equation}
\label{eq:CharacteristicFctWTDlimit1}
    \frac{\mathcal{Z}_{\Nc,2}(p)}{\mathcal{Z}_{\Nc,2}(0)} \underset{p\to0}{\simeq} 1 - B_\Nc\,\sqrt{p}
 \:,
\end{equation}
where $B_\Nc$ is some constant. This behaviour can now be related to the distribution by using a Tauberian theorem. Assuming the tail $\Rdwt{\Nc}{2}(t)\simeq c\,\rt^{-3/2}$, we have
\begin{eqnarray}
     \hspace{-2cm}
   \int_0^\infty\D\rt\,
  \Rdwt{\Nc}{2}(t)\, \EXP{-\Nc p\rt}
  =1 - \int_0^\infty\D\rt\,
  \Rdwt{\Nc}{2}(t)\, (1- \EXP{-\Nc p\rt} )
     \\
     \nonumber
     \hspace{-2cm}
   \underset{p\to0}{\simeq }
   1 - c \int_0^\infty\frac{\D\rt}{\rt^{3/2}}\, (1- \EXP{-\Nc p\rt} )
   = 1 -   2\,c \,\Nc\,p \int_0^\infty\D\rt\,\rt^{-1/2}\, \EXP{-\Nc p\rt}
   = 1 - 2\,c \,\sqrt{\pi\,\Nc\,p }
\end{eqnarray}
Thus $c=B_\Nc/(2\sqrt{\pi\Nc})$.
A precise determination of $B_\Nc$ would be interesting, in particular in order to clarify the precise scaling with $\Nc$ of the typical values of the random variable $\Wt$, however it goes beyond the present analysis.

We conclude that in the limit of small transmission, $\coupl\ll1$, the Wigner time delay distribution shows the universal $\tau^{-3/2}$ behaviour
\begin{equation}
  \label{eq:UniversalTailDistWTD}
  \dwt{\Nc}{2}(\tau)  \sim
  \coupl^{-1}\,(\coupl/\tau)^{3/2}
  \hspace{1cm}\mbox{for }
  \coupl \ll \tau \ll 1/\coupl
  \:.
\end{equation}
In the next section, we will see that the upper cutoff also carries a $\Nc$-dependence. This behaviour coincides with the one obtained by a heuristic argument, see Eq.~\eref{eq:heuristic1}, which is based on the picture of isolated resonances.

It is worth stressing that the order of the limits $p\to0$ and $\fss\to0$ is important. For finite coupling the first moments are finite. Using that the second moment is \cite{LehSavSokSom95,FyoSom97} $\mean{\Wt^2}\simeq1/(2\fss\Nc^3)$, cf. Eq.~\eref{eq:FyodorovSommers1997Eq195}, one expects
\begin{equation}
    \frac{\mathcal{Z}_{\Nc,\beta}(p)}{\mathcal{Z}_{\Nc,\beta}(0)} \underset{p\to0}{\simeq}
    1  - \frac{2p}{\beta\fss} + \frac{p^2}{\beta^2\fss^3\Nc} +\cdots
\end{equation}
for small but finite $\fss$ (and large $\Nc$). The behaviour \eref{eq:CharacteristicFctWTDlimit1} is obtained by sending first $\fss\to0$ and \textit{then} $p\to0$. For finite $\fss$ the non-analyticity of the characteristic function appears at higher order in $p$, corresponding to the divergence of the moments of high order, $\smean{\Wt^k}=\infty$ for $k\geq1+\beta\Nc/2$.

For finite $\coupl$, the distribution $\dwt{\Nc}{\beta}(\tau)$ should be in correspondence with the marginal distribution of the proper (or partial) times in the limit $\tau\to\infty$, with $\mprop{\Nc}{\beta}(\tau)\sim\tau^{-2-\beta\Nc/2}$, as we expect that one proper time dominates the sum $\Wt=(1/\Nc)\sum_a\tau_a$.
Inspection of the matrix distribution \eref{eq:DistributionZGeneral} shows that if one resonance is much
more narrow than all others,  $\gamma_1\to0$, we expect the vanishing of the density as
$P_\invQ(\Gamma)\sim(\det\invQ)^{\beta\Nc/2}\sim\gamma_1^{\beta\Nc/2}$.
Correspondingly the distribution of $\Wt=(1/\Nc)\sum_a\gamma_a^{-1}\simeq1/(\Nc\,\gamma_1)$ presents the tail $\dwt{\Nc}{\beta}(\tau) \sim \tau^{-2-\beta\Nc/2}$.
We can reintroduce the dependence in $\coupl$ by matching the behaviour with \eref{eq:UniversalTailDistWTD}~:
\begin{equation}
  \dwt{\Nc}{\beta}(\tau) \underset{\tau\gg1/\coupl}{\sim }
  \coupl^2\,(\coupl\tau)^{-2-\beta\Nc/2}
  \:,
\end{equation}
for $\tau\gtrsim1/\coupl$.
A similar decoupling of the eigenvalues was demonstrated for perfect contacts in Ref.~\cite{TexMaj13}. Note that the $\Nc$-dependence has not been included above. This will be discussed in the Section~\ref{sec:Numerics} (see also section~\ref{Subsec:Heuristic}, where such a behaviour has been related to isolated resonances with atypically narrow width).

\subsection{Large deviations for $\tau\to0$ for arbitrary symmetry class}

In this last subsection, we study the limiting behaviour of the distribution $\dwt{\Nc}{\beta}(\tau)$ for $\tau\ll\fss/\Nc$ by a steepest descent analysis of the matrix integral, which allows to consider any symmetry class.
Our starting point is
\begin{eqnarray}
\hspace{-2cm}
   \mathcal{Z}_{\Nc,\beta}(p) \propto
    \int_{\invQ>0}\mathrm{D}\invQ\, (\det \invQ )^{\beta\Nc/2}
    &\int\mathrm{D}\Kmat\,
    \frac{ \det(\identity_\Nc+\Kmat^2)^{ \beta\Nc/2 } }
         { \det(\identity_\Nc+\fss^2\Kmat^2)^{\beta\Nc+1-\beta/2} }
         \\
         \nonumber
         &\times
   \exp\left(
     - \frac{\beta}{2}\fss\, \tr{ \frac{\identity_\Nc+\Kmat^2}{\identity_\Nc+\fss^2\Kmat^2} \invQ }
     - \frac{2p}{\beta\fss}\,\tr{\invQ^{-1}}
     \right)
     \:.
\end{eqnarray}
The integral over the matrix $\invQ$ is of the form of the Bessel function with matrix argument introduced in Ref.~\cite{Her55}, generalising the MacDonald function as
\begin{equation}
  \label{eq:MatrixMacDonald}
   B_{\nu,\beta}(Z)
   = \int_{X>0}\mathrm{D}X\,
   \left(\det X\right)^{-\nu-1-\beta(N-1)/2} \,\EXP{-\tr{X+Z\,X^{-1}}}
   \:,
\end{equation}
where $Z$ is a Hermitian matrix.
The relation with the characteristic function reads explicitly
\begin{eqnarray}
 \label{eq:RepresZMartrixBessel}
  \hspace{-1cm}
   \mathcal{Z}_{\Nc,\beta}(p) \propto
   \int\mathrm{D}\Kmat\,
    \frac{ \det(\identity_\Nc+\Kmat^2)^{ \beta\Nc/2 } }
         { \det(\identity_\Nc+\fss^2\Kmat^2)^{\beta\Nc+1-\beta/2} }
   \:
   B_{1+\frac{\beta\Nc}{2},\beta}\left(p\,\frac{\identity_\Nc+\Kmat^2}{\identity_\Nc+\fss^2\Kmat^2}\right)
   \:.
\end{eqnarray}

The limiting behaviour of integrals such as \eref{eq:MatrixMacDonald} was recently studied by the Laplace method in \cite{ButWoo03} for real symmetric matrices.
Here we generalise this analysis to the unitary class, which allows us to compute easily the remaining matrix integral (over $\Kmat$). Using the invariance under unitary transformations, we can always choose one of the two matrices under a diagonal form.
We choose $\Kmat=\mathrm{diag}(k_1,\cdots,k_\Nc)$.
Next we perform the change of variable
$$
\invQ\longrightarrow
\frac{2\sqrt{p}}{\beta\fss}
\left(\frac{\identity_\Nc+\Kmat^2}{\identity_\Nc+\fss^2\Kmat^2}\right)^{-1/4}
X
\left(\frac{\identity_\Nc+\Kmat^2}{\identity_\Nc+\fss^2\Kmat^2}\right)^{-1/4}
$$
Thus
\begin{eqnarray}
\label{eq:84}
\hspace{-2cm}
   \mathcal{Z}_{\Nc,\beta}(p) \propto
   p^{\frac{\beta\Nc^2}{2}+\frac{\Nc}{2}(1-\beta/2)}
   \int\D k_1\cdots\D k_\Nc\,|\Delta_\Nc(k)|^\beta
   \prod_n
   \frac{(1+k_n^2)^{(\beta/2-1)/2}}{(1+\fss^2k_n^2)^{\beta\Nc/2-\beta/4+1/2}}
   \nonumber\\
   \times
   \int_{X>0}\mathrm{D}X\,  (\det X )^{\beta\Nc/2}
   \exp\left(
     - \sqrt{p}\, \tr{ \sqrt{\frac{\identity_\Nc+\Kmat^2}{\identity_\Nc+\fss^2\Kmat^2}} \left(X+X^{-1}\right)}
     \right)
     \:.
\end{eqnarray}
Then, we introduce
$R=(\identity_\Nc+\Kmat^2)^{1/2}(\identity_\Nc+\fss^2\Kmat^2)^{-1/2}$.
The integral is dominated by the position of the saddle point, minimum of $\tr{R\,(X+X^{-1})}$, which is found to be $X_*=\identity_\Nc$.
The Hessian matrix has the form
$\mathscr{H}_{(i,j),(k,l)}=2(R_{jl}\delta_{ik}+R_{ik}\delta_{jl})$, so that we obtain the form
\begin{eqnarray}
   \int_{X>0}\mathrm{D}X\,  (\det X )^{\beta\Nc/2}
   \EXP{ - \Lambda\, \tr{ R\, \left(X+X^{-1}\right)}   }
     \nonumber
     \\
     \underset{\Lambda\to\infty}\simeq
     \left(\frac{\pi}{\Lambda}\right)^{\Nc(1+\beta(\Nc-1)/2)}
     (\det R)^{-1/2}
       \prod_{i<j}(R_{ii}+R_{jj})^{-\beta/2}
       \,\EXP{-2\Lambda\,\tr{ R }}
\end{eqnarray}

After some algebra we eventually get the limiting behaviour (assuming $\fss\to0$)
\begin{equation}
  \mathcal{Z}_{\Nc,\beta}(p) \propto
  p^{\beta\Nc^2/4}\,\EXP{-2\Nc\sqrt{p}}
  \hspace{1cm}
  \mbox{for }
  p\to\infty
\end{equation}
which agrees with \eref{eq:CaraFctWTDBeta2} for $\beta=2$.
Correspondingly, we obtain the limiting behaviour for the distribution of the Wigner time delay
\begin{equation}
  \label{eq:LargeDevSmallTauWC}
  \dwt{\Nc}{\beta}(\tau)
  \sim
  \tau^{-\frac{\beta\Nc^2}{2}-\frac32}\,
  \EXP{-\beta\Nc\fss/(2\tau)}
  \hspace{1cm}
  \mbox{for }
  \tau\to0
  \mbox{ and }\fss\ll1
  \:.
\end{equation}
As a check, we can compare this behaviour with the limiting behaviour of the marginal distribution for proper times and partial times as the three distributions coincide for one channel, $\dwt{1}{\beta}(\tau)=\mprop{1}{\beta}(\tau)=\mpart{1}{\beta}(\tau)$.
From Eq.~\eref{eq:LimitsMarginalPartialTimes} we have
$\mpart{\Nc}{\beta}(\tau)\sim\tau^{-\beta\Nc/2-3/2}\,\exp\big\{-\beta\fss/(2\tau)\big\}$
and from Eq.~\eref{eq:LimitsMarginalProperTimes},
$\mprop{\Nc}{2}(\tau)\sim\tau^{-2\Nc-1/2}\,\exp\big\{-\fss/\tau\big\}$.
The three limiting behaviours indeed coincide when $\Nc=1$, as it should.

For reference, we can compare this behaviour to the corresponding one for ideal couplings~(see Ref.~\cite{TexMaj13} and Section~5 of Ref.~\cite{GraTex16b}, and also \cite{Gra18})
\begin{equation}
  \label{eq:LargeDevSmallTauPC}
  \dwt{\Nc}{\beta}^{(0)}(\tau)
  \sim
  \tau^{-\frac{3\beta\Nc^2}{4}-\frac{\Nc}{2}(1-\frac{\beta}{2})-\frac32}\,
  \EXP{-\beta\Nc/(2\tau)}
  \hspace{1cm}
  \mbox{for }
  \tau\to0
  \mbox{ and  }\fss=1
  \:.
\end{equation}
Although the leading exponential terms in \eref{eq:LargeDevSmallTauWC} and \eref{eq:LargeDevSmallTauPC} coincide, the pre-exponential factors there have different power law dependencies.


\section{Numerical analysis}
\label{sec:Numerics}

We have performed numerical simulations in order to study the weak coupling limit.
For this purpose we use the formulation presented in Section~\ref{sec:IntegrationOverCUE}~:
we generate the matrix $\Sm_0$ in the circular ensemble and the matrix $\invQ_0=\WSm_{s0}^{-1}$ in the Laguerre ensemble. The Wigner-Smith matrix is then constructed making use of the expression
\begin{equation}
  \WSm = \left( 1-|\Sbar|^2 \right) \,
  \left( \identity_\Nc + \Sbar\Sm_0^\dagger  \right)^{-1} \, \Sm_0^{-1/2}
  \WSm_{s0}\,
  \, \Sm_0^{1/2}
  \left( \identity_\Nc + \Sbar^*\Sm_0  \right)^{-1}
  \:.
\end{equation}

\begin{figure}[!ht]
\centering
\includegraphics[height=6cm]{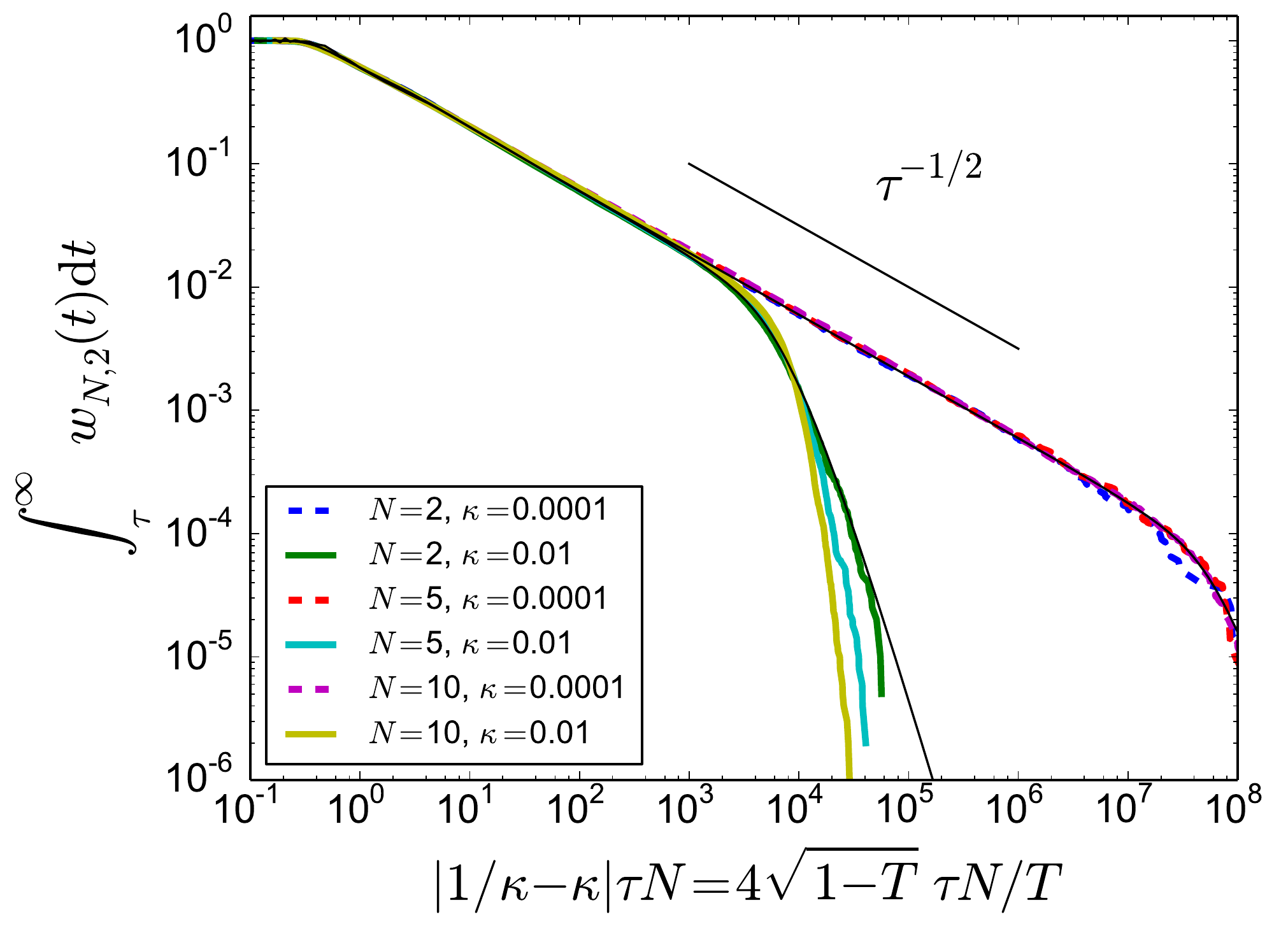}
\caption{\it
  Cumulative distribution of the proper time delays in the unitary case ($\beta=2$) for different channel number $\Nc$ and coupling $\fss$ (the latter controls the transmission probability through the contact, $\coupl\simeq4\fss$ at small $\fss\ll1$). The dashed black line corresponds to the exact analytical expression~\eref{eq:SommersSavinSokolov2001a}.
  }
\label{fig:CumulativeProperTimes}
\end{figure}

\subsection{Check: marginal distribution of the proper times}

As a first check, we have computed the cumulative (marginal) distribution of the proper time for different $\Nc$ and $\fss$. This distribution is shown in Fig.~\ref{fig:CumulativeProperTimes}, where it is plotted in terms of the scaling variable $s=\Nc|1/\fss-\fss|\,\tau$, which is a natural choice describing the full range of couplings (see \ref{app:PartialProper}). We have generated $10^{5}$ matrices each time.
For weak coupling $\fss\to0$, the main behaviours of the distribution are
\begin{equation}
  \frac{\fss}{\Nc}\,\mprop{\Nc}{\beta}\left(\tau=\frac{\fss}{\Nc}\,s\right)
  \sim
  \left\{
  \begin{array}{ll}
  s^{-3/2}   &  \mbox{for } 1 \lesssim s \lesssim 1/\fss^2
  \\[0.25cm]
  \fss^3 \left(\fss^2 s \right)^{-2-\beta\Nc/2}  &  \mbox{for } s \gtrsim 1/\fss^2  \\
  \end{array}
  \right.\,,
\end{equation}
which are deduced in \ref{app:PartialProper} from the known exact result~\cite{SomSavSok01}. We can see that in the limit $\fss\to0$ all curves collapse onto each other (after proper rescaling). Changing $\fss$ then only shifts the upper cutoff of the $s^{-3/2}$ tail. The positions of the lower and upper cutoffs of this power law perfectly coincide with the two cutoffs $\tup$ and $\tlow$ defined by Eqs.~(\ref{eq:DefUpperCutoffProper}) and (\ref{eq:DefLowerCutoffProper}).
We have also compared the numerics with the exact distribution \eref{eq:SommersSavinSokolov2001a} for $\Nc=2$ (in practice, this is only possible for small $\Nc\lesssim5$ and not too small $\fss\gtrsim0.01$, otherwise \eref{eq:SommersSavinSokolov2001a} appears to be too involved for being plotted with a conventional software like {\tt Mathematica}): the agreement is excellent.

\begin{figure}[!ht]
\centering
\includegraphics[width=0.6\textwidth]{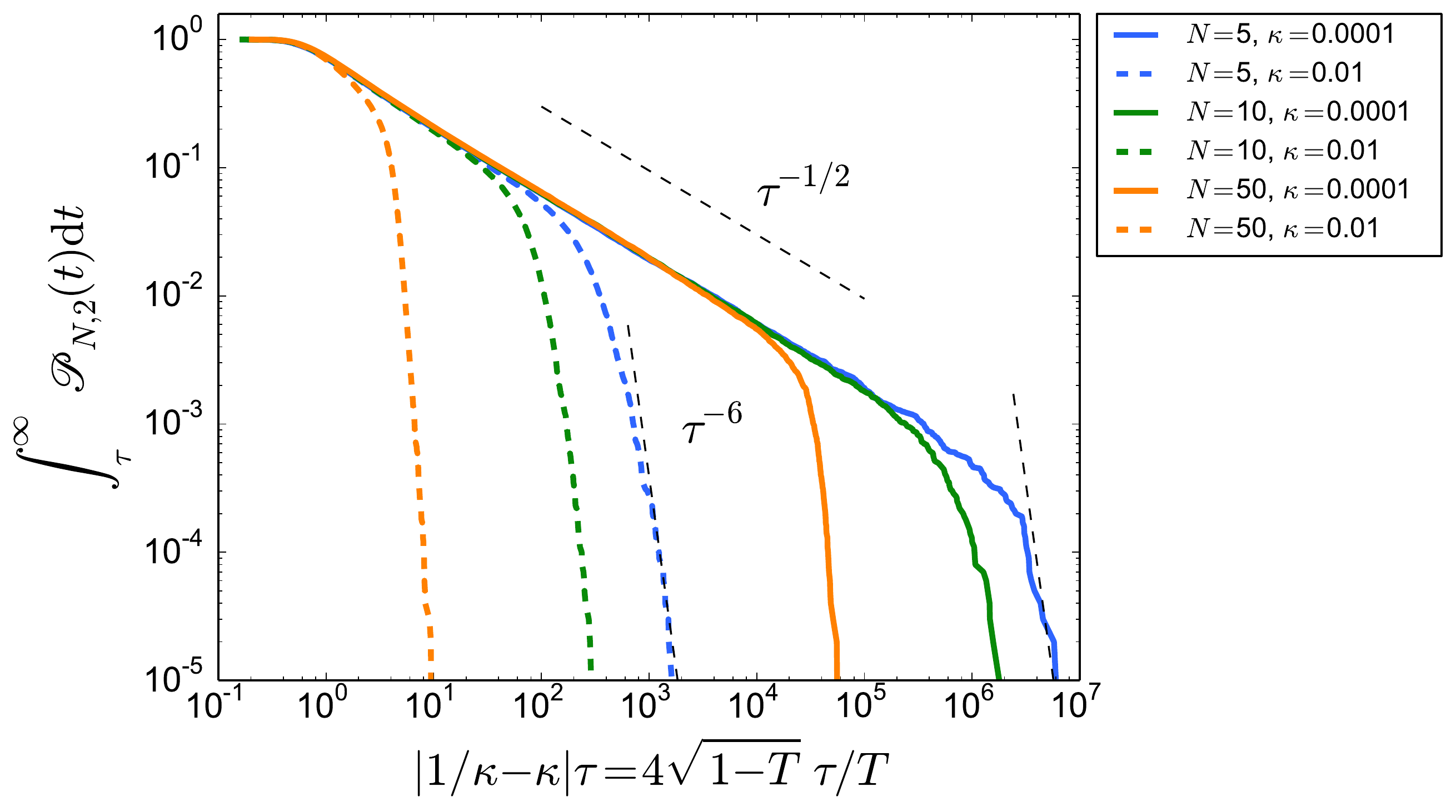}
\includegraphics[width=0.49\textwidth]{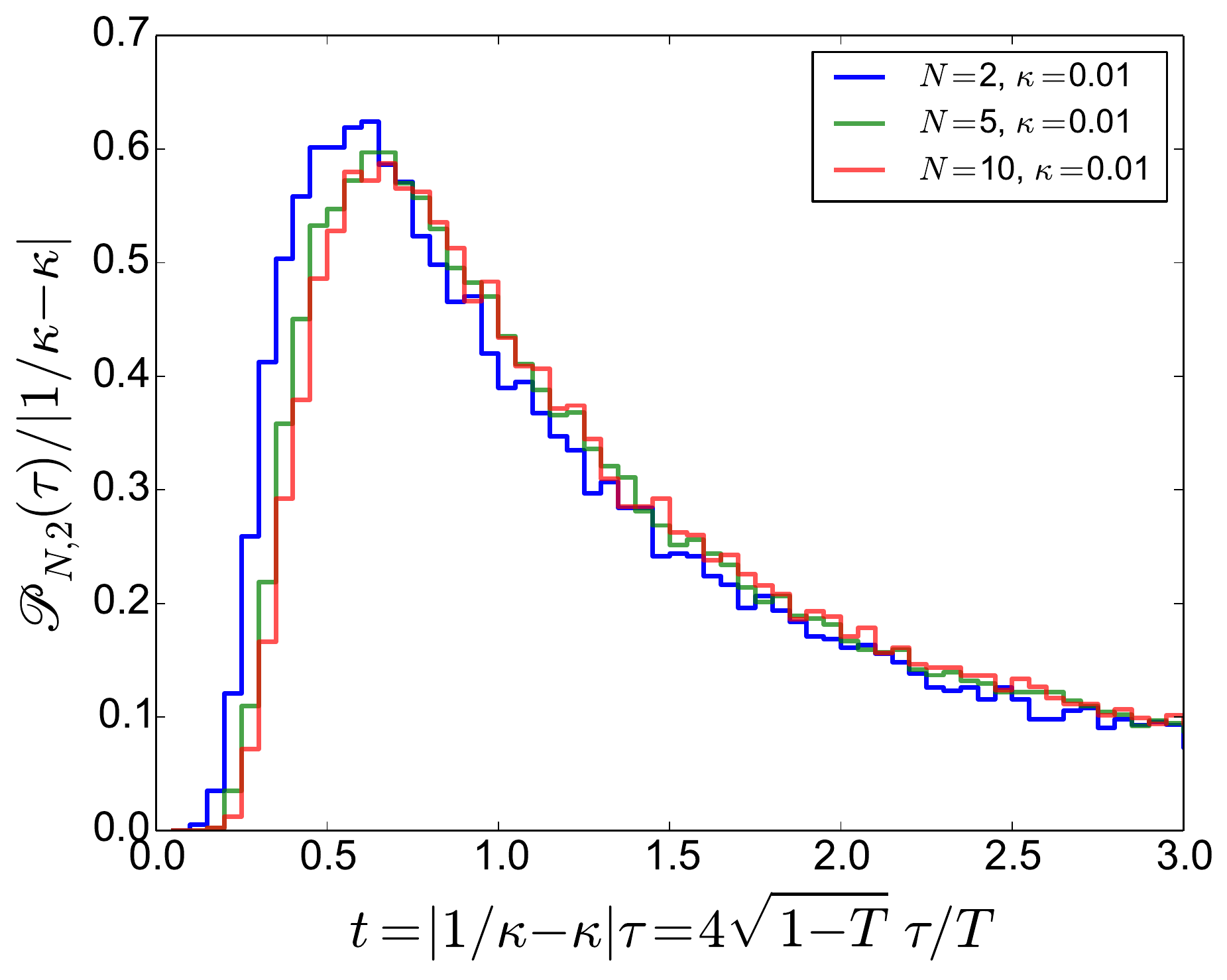}
\caption{\it Cumulative distribution of the Wigner time delay in the unitary case ($\beta=2$) for different channel numbers and different couplings.
The dashed black lines are $\tau^{-1/2}$ and $\tau^{-1-\Nc}$.
The distributions for different channel numbers are plotted for $\fss=0.01$ on the bottom part of the figure.
}
\label{fig:CumulativeWignerTime}
\end{figure}

\subsection{Distribution of the Wigner time delay}

Next, we have considered the distribution of the Wigner time delay in the weakly coupled regime, $\Nc\coupl\ll1$. We draw several conclusions from such a numerical analysis.

\begin{figure}[!ht]
\centering
\includegraphics[width=0.55\textwidth]{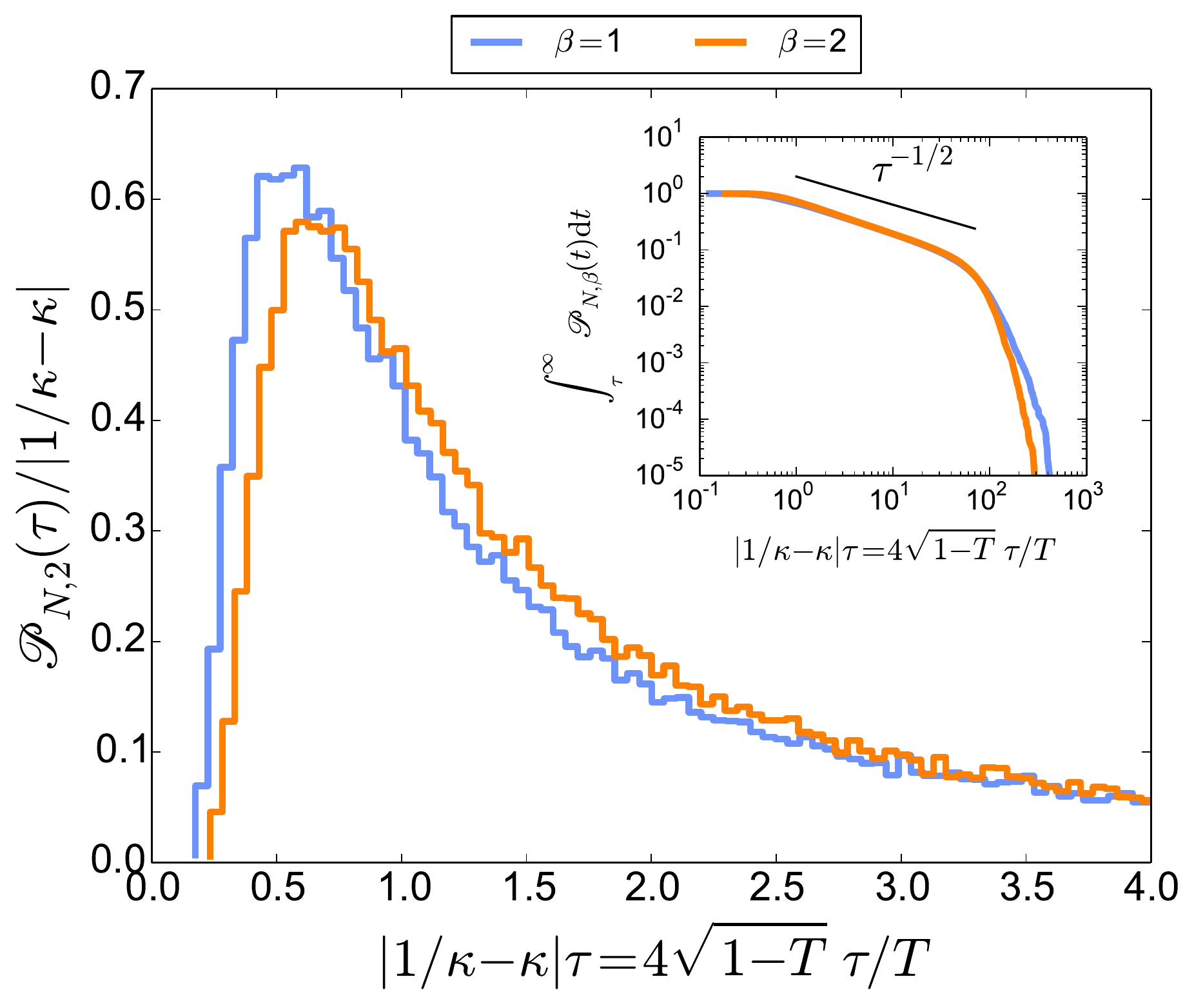}
\caption{\it Comparison of the cumulative distribution of the Wigner time delay in the orthogonal and unitary case.}
\label{fig:ComparisonOrthoUnitary}
\end{figure}

\begin{itemize}
\item
  Taking again $s=|1/\fss-\fss|\,\tau$ as the scaling variable, we see that the different distributions collapse onto each other and show the intermediate $s^{-3/2}$ behaviour for different $\Nc$ and $\fss$ (Fig.~\ref{fig:CumulativeWignerTime}).
\item
   The lower cutoff of the $s^{-3/2}$ law is almost independent of $\Nc$.
\item
  The upper cutoff depends on both $\fss$ and $\Nc$, with numerics supporting the scaling $\tau_*\sim1/(\fss\Nc^2)$. (This can be clearly seen, e.g., by comparing the two curves for $\Nc=5$ and $\Nc=50$ in Fig.~\ref{fig:CumulativeWignerTime} for the same value of $\fss$.)
\item
  The power law $\tau^{-3/2}$ is observed both in the unitary and orthogonal case (Fig.~\ref{fig:ComparisonOrthoUnitary}). (This is consistent with the earlier study~\cite{FyoSavSom97} of the crossover regime).
\item
  For $\tau\gtrsim\tau_*$, the distribution exhibits a power law tail with the universal exponent $2+\beta\Nc/2$, which is anticipated theoretically and confirmed here numerically.
\end{itemize}
These findings together with the outcome of Section~\ref{sec:characfct} can be summarised as follows~:
\begin{equation}
  \fss\,  \dwt{\Nc}{\beta}\left(\tau=\fss\,s\right)
  \sim
  \left\{
  \begin{array}{ll}
  s^{-3/2}
  &  \mbox{for } 1 \lesssim s \lesssim 1/(\fss\Nc)^2
  \\[0.25cm]
  (\Nc\fss)^3 \left(\Nc^2\fss^2s\right)^{-2-\beta\Nc/2}
  &  \mbox{for } s \gtrsim 1/(\fss\Nc)^2  \\
  \end{array}
  \right.
\end{equation}
Furthremore, we have argued in Section~\ref{subsec:UniversalLimitWC} that
$\lim_{\fss\to0}\fss\,  \dwt{\Nc}{\beta}\left(\tau=\fss\,s\right)$ is a universal function, although we have not been able to determine its precise form.

\begin{figure}[!ht]
\centering
\includegraphics[width=0.6\textwidth]{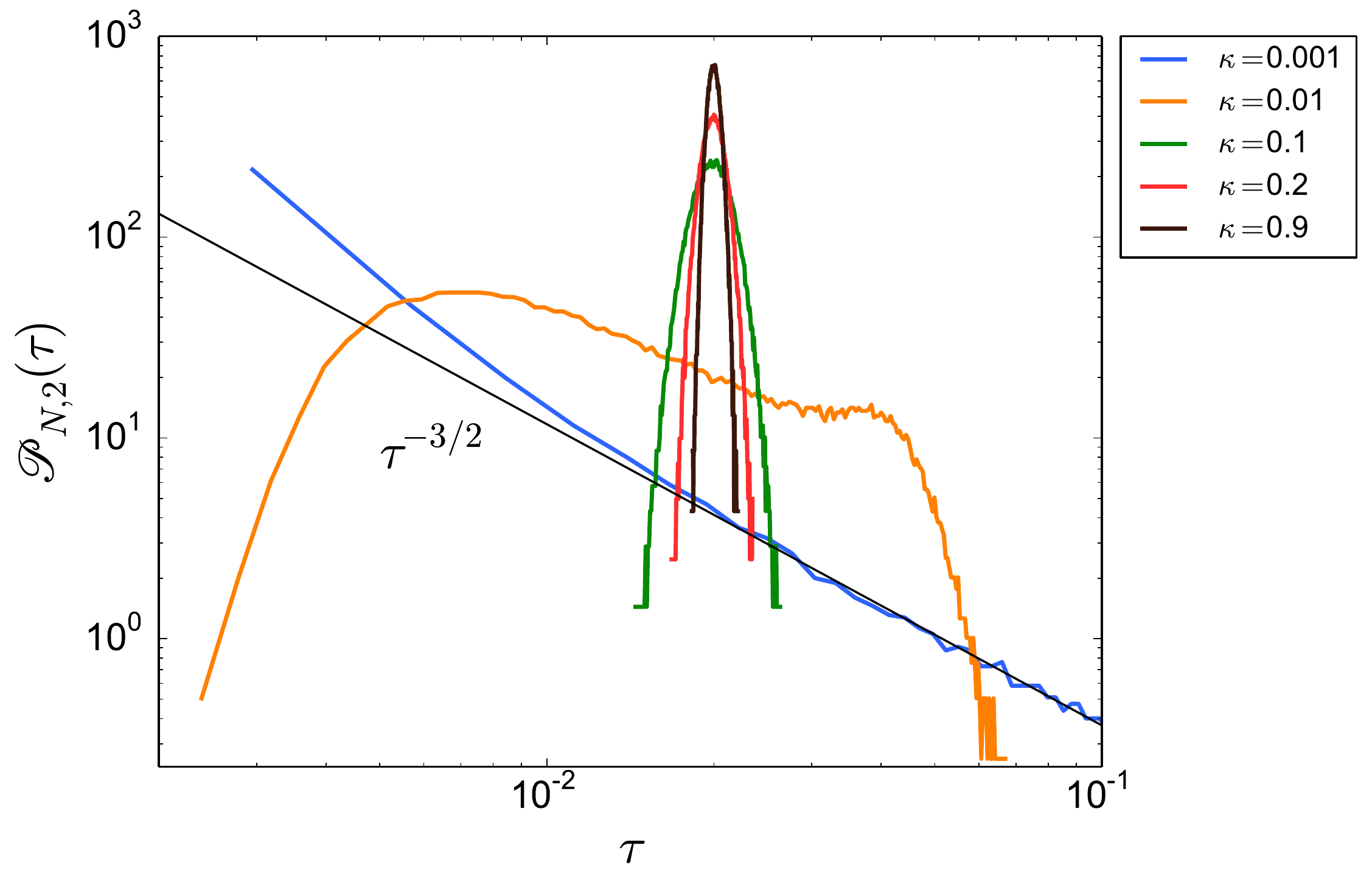}
\caption{\it Distribution of the Wigner time delay for $\Nc=50$ channels~: from strong coupling regime ($\Nc\coupl\gg1$) to weak coupling  ($\Nc\coupl\ll1$).}
\label{fig:DWTfromSCtoWC}
\end{figure}

Finally, we have also studied the transition from strong coupling ($\Nc\coupl\gg1$) to weak coupling  ($\Nc\coupl\ll1$), for large $\Nc$, and shown that the distribution crosses over from a narrow distribution to a broad distribution when $\Nc\coupl\sim1$ (Fig.~\ref{fig:DWTfromSCtoWC}).

\section{Conclusion}

In this article, we have considered the scattering of waves by a chaotic cavity
coupled to $\Nc$ channels characterised by arbitrary transmission coefficients $T$. Within a random matrix approach, we have derived the joint distribution of the scattering matrix $\Sm$ and the symmetrised time-delay matrix $\WSm_s$ at arbitrary channel couplings. This extends the result obtained by Brouwer, Frahm and Beenakker \cite{BroFraBee97,BroFraBee99} at $\coupl=1$ to the general case of non-ideal coupling, $\coupl<1$. This has allowed us to obtain two representations for the distribution of $\WSm_s$ (or more precisely, its inverse) in terms of certain matrix integrals.

Then we have applied our results to study the statistical properties of the Wigner time delay $\Wt=\frac{1}{\Nc}\tr{\WSm_s}$. Specifically, we have derived the exact representation \eref{eq:RepresZMartrixBessel} of the characteristic function of $\Wt$ as a multiple integral involving Bessel functions of matrix argument. This expression has been further used to obtain, after inverse Laplace transform, the asymptotic behaviours of the Wigner time delay distribution in the limit $\coupl\to0$ (weak coupling per channel), keeping $\Nc\coupl\ll1$. Physically, this corresponds to the regime of isolated resonances (the system weakly coupled to the external). In such a case, the Wigner time delay distribution becomes broad with an intermediate behaviour described by the universal $\tau^{-3/2}$ law. We have also established the left and right tails of the distribution up to the constant prefactors that have not been computed.
The knowledge of these constants would however be needed for determining the precise positions of the crossovers between the limiting behaviours. These cutoffs are of interest, as they control the positive and negative moments, but they have only been deduced here from a numerical analysis.
We have also compared such a behaviour with the one derived from the known
marginal distributions of the partial and proper time delays, which become almost identical to each other in the weak coupling limit (see \ref{app:PartialProper}). In particular, the distribution of the partial time delays (rescaled properly by $T$) is found to have a simple universal form \eref{eq:MarginalPartialTimeGoparMelloMethod} in the limit $\coupl\to0$ at any $\Nc$.
We have argued that the distribution of the Wigner time delay should be described by a universal function in the limit $\coupl\to0$ as well. The analysis of the other
regime with $\Nc\coupl\gg1$ (the strongly coupled system with overlapping resonances) suggests that the Wigner time delay distribution becomes narrow, with a Gaussian-like bulk behaviour. The crossover between the two limiting forms occurs quite sharply at $\Nc\coupl\sim1$ (cf. Fig.~\ref{fig:DWTfromSCtoWC}). Determining the precise universal function describing such a crossover is still an outstanding problem and a challenging one to consider in future study.


\section*{Acknowledgements}

We thank Pierpaolo Vivo for stimulating discussions.
DVS gratefully acknowledges University Paris-Sud for financial support and LPTMS in Orsay for hospitality during his stay there.


\begin{appendix}


\section{Some matrix integrals}

\subsection{Harish-Chandra - Itzykson - Zuber Integrals}
\label{app:HCIZ}

Consider two Hermitian matrices $A$ and $B$ with spectra $\{a_i\}$ and $\{b_i\}$.
Then \cite{ItzZub80a,ZinZub03}
\begin{equation}
  \hspace{-1cm}
  \int_{\mathrm{U}(N)} \mathrm{D}U\,
  \exp\left(t\,\tr{ AUBU^\dagger} \right)
  =
  G(N+1)\,
  t^{-N(N-1)/2}
  \frac{\det\left(\EXP{t\,a_ib_j}\right)_{1\leq i,\,j\leq N}}{\Delta_\Nc(a)\,\Delta_\Nc(b)}
\end{equation}
where
\begin{equation}
  \Delta_\Nc(a) = \det(a_i^{j-1})_{1\leq i,\,j\leq N} =\prod_{i<j}(a_i-a_j)
\end{equation}
is the Vandermonde and $G(z)$ is the Barnes' $G$ function (double gamma function) \cite[\S5.17]{DLMF} defined by $G(z+1)=\Gamma(z)G(z)$, i.e. $G(N+1)=(N-1)!(N-2)!\cdots3!2!1!$.


\subsection{Two normalisation constants}
\label{app:tmi}

We state two matrix integrals provided in Forrester's book~\cite{For10}, which are used in the article. The normalisation for the Cauchy ensemble is given by Eqs.~(4.4) and (4.145) of \cite{For10}:
\begin{equation}
  \label{eq:mi1}
  \hspace{-2cm}
  \int\D x_1\cdots\D x_N\,|\Delta_N(x)|^\beta\,
   \prod_n (1+x_n^2)^{-\alpha}
  =2^{\beta N(N-1)/2-2(\alpha-1)N} \pi^N\, M_N(a,a,\beta/2)
\end{equation}
where $a=\alpha-1-\beta(N-1)/2$ and
\begin{equation}
      M_N(a,b,\lambda)
    =\frac{1}{\Gamma(1+\lambda)^N}\prod_{j=0}^{N-1}
    \frac{\Gamma(\lambda j +a+b+1) \Gamma(\lambda(j+1)+1)}
           {\Gamma(\lambda j +a +1)\Gamma(\lambda j  +b+1)}
           \:.
\end{equation}
The normalisation for the Gaussian ensemble is given on p.~173 of~\cite{For10}:
\begin{equation}
   \label{eq:mi2}
   \int\D x_1\cdots\D x_N\,|\Delta_N(x)|^\beta\,
   \prod_n \EXP{-x_n^2/2}
   = \frac{(2\pi)^{N/2}}{\Gamma(1+\beta/2)^{N}}\,
   \prod_{j=1}^{N}\Gamma(1+j\beta/2)
   \:.
  \end{equation}


\section{Andr\'eief formula}
\label{app:Andreief}

A formula due to Andr\'eief \cite{And86} (see also the recent historical note \cite{For18}) is
\begin{equation}
  \hspace{-1.5cm}
  \int \left(\prod_{n=1}^N\D\mu(x_n)\right)
  \det(A_i(x_j))\, \det(B_k(x_l))\,
  =N!\, \det\left[\int\D\mu(x)\,A_i(x)\,B_j(x)\right].
\end{equation}

For $\beta=2$, writing the Vandermonde as
\begin{equation}
  \Delta_\Nc(\lambda)^2=
  \prod_{i<j}(\lambda_i-\lambda_j)^2
  =
  \underbrace{ \det(\lambda_i^{k-1}) }_{\prod_{i<j}(\lambda_i-\lambda_j)}
  \det(\lambda_j^{k-1})
\end{equation}
we deduce the representation of the matrix integral as a Hankel determinant
\begin{equation}
  \int \left(\prod_{i=1}^N\D\mu(\lambda_i)\right)
  \prod_{i<j}(\lambda_i-\lambda_j)^2
  =N!\: \det\left( a_{ij} \right)_{1\leq i,\,j\leq N}
  \:,
\end{equation}
where the matrix elements are
\begin{equation}
  a_{ij} = \int\D\mu(\lambda)\,\lambda^{i+j-2}
  \hspace{1cm}\mbox{for } 1\leq i,\:j\leq N
  \:.
\end{equation}


\section{Partial and proper time delays}
\label{app:PartialProper}

The marginal distributions of the partial and  proper time delays were obtained in several papers by Fyodorov, Sommers and collaborators \cite{FyoSom96,FyoSom97,FyoSavSom97,SavFyoSom01} (partial times) and \cite{SomSavSok01} (proper times). These explicit results are however expressed in complicated forms, with the transmission coefficient entering through the following parameter~:
\begin{equation}
  g  = \frac{2}{\coupl} - 1 = \frac{1}{2}\left( \fss  +\frac{1}{\fss} \right) \geq1\,.
\end{equation}
It is the purpose of this appendix to derive the precise limiting behaviours of these distributions in the weak coupling limit $\coupl\approx2/g\to0$. It will be convenient to rescale the time delays and relevant distributions as follows
\begin{equation}
 \label{eq:rescaling}
 \hspace{-1cm}
  \tau \simeq \frac{\beta}{4g}\,\rt
  \simeq \frac{\beta\fss}{2}\,\rt
  \simeq  \frac{\beta\coupl}{8}\,\rt
  \hspace{1cm}\mbox{and} \hspace{1cm}
  \Rmprop{\Nc}{\beta}(\rt)
  \underset{g\gg1}{\simeq   }
  \frac{\beta}{4g}\,
  \mprop{\Nc}{\beta}\left(\tau\simeq\frac{\beta}{4g}\rt \right)
  \:,
\end{equation}
with a similar form for $\Rmpart{\Nc}{\beta}(\tau)$.

\subsection{Marginal distribution of the partial time delays in the unitary case}

The marginal distribution of partial time delays was first derived by Fyodorov and Sommers~\cite{FyoSom96,FyoSom97} in the unitary case:
\begin{equation}
  \hspace{-1cm}
  \mpart{\Nc}{2}(\tau) = \frac{1}{\tau^2}\,\tilde{p}_\Nc^{(2)}(1/\tau)\,,
  \hspace{0.5cm}\mbox{where }
  \tilde{p}_{\Nc}(\gamma) = \frac{\gamma^\Nc}{\Nc!}\left(-\partial_\gamma\right)^\Nc
  \left[
    I_0(\sqrt{g^2-1}\,\gamma)\EXP{-g\gamma}
  \right]
  \:.
\end{equation}
In order to find limiting behaviours we rescale the distribution by introducing $\rt=2g\tau$ or $z=\gamma/(2g)$.

We first consider the domain $z\ll1$ (i.e. $\tau\ll g$).
Using that
\begin{equation}
  I_0(\sqrt{g^2-1}\,\gamma)\EXP{-g\gamma} \simeq \frac{1}{2g}\,\phi\left( z=\frac{\gamma}{2g}\right)
  \hspace{1cm}\mbox{with }
  \phi(z) \eqdef \frac{1}{\sqrt{\pi z}}\EXP{-z}
\end{equation}
we write
\begin{equation}
  \tilde\pi_{\Nc}(z) =\lim_{g\to\infty}  2g \:\tilde{p}_{\Nc}(\gamma=2g\,z)
  = \frac{z^\Nc}{\Nc!}\left(-\partial_z\right)^\Nc
  \left[
    \phi(z)
  \right]
\end{equation}
Using $(-\partial_z)^n(1/\sqrt{z})=(1/2)_n z^{-1/2-n}=2^{-n}(2n-1)!!\,z^{-1/2-n}$, where $(a)_n=a(a+1)\cdots(a+n-1)=\Gamma(a+n)/\Gamma(a)$ is the Pochhammer symbol, we deduce
\begin{equation}
  \tilde\pi_{\Nc}(z) = \frac{1}{\Nc!}\left(
    \sum_{n=0}^\Nc C_\Nc^n \,\frac{(2n-1)!!}{2^n}\,z^{\Nc-n}
  \right) \,\phi(z)
  \:.
\end{equation}
We obtain
\begin{equation}
  \label{eq:MarginalPartialTimesWC}
  \lim_{\fss\to0} \Rmpart{\Nc}{2}(\rt)
  =   \frac{\EXP{-1/\rt}}{\sqrt{\pi}\,\rt^{3/2}}\,
  \frac{1}{\Nc!}\sum_{n=0}^\Nc \frac{C_\Nc^n(2\Nc-2n-1)!!}{2^{\Nc-n}} \, \rt^{-n}
\end{equation}
In particular, for $t\gg1$, we get
\begin{equation}
  \label{eq:FyodorovSommersLimiting1}
  \lim_{\fss\to0} \Rmpart{\Nc}{2}(\rt) \simeq \frac{(2\Nc-1)!!}{\sqrt{\pi}\,2^\Nc\,\Nc!}\,\rt^{-3/2}
  \:.
\end{equation}

We now turn to the study of the far tail ($\tau\gg1/\fss$).
We expand $I_0(\sqrt{g^2-1}\gamma)\EXP{-g\gamma}$ in powers of $\gamma$ and identify the coefficient of the term $\gamma^\Nc$.
Some algebra gives the form
\begin{equation}
  \label{eq:FyodorovSommersLimiting2}
  \Rmpart{\Nc}{2}(\rt)
  \simeq  \frac{a_\Nc}{g^3} \, (\rt/g)^{-2-\Nc}
\end{equation}
where
\begin{equation}
\hspace{-1cm}
  a_\Nc = 2 \sum_{m=0}^{\lfloor\Nc/2\rfloor}
  \frac{2^{\Nc-2m}}{(m!)^2(\Nc-2m)!}
  = \frac{2^{1+2\Nc}\,\Gamma(\frac{1}{2}+\Nc)}{\sqrt{\pi}\,(\Nc!)^2}
  = \frac{2^{1+\Nc}\,(2\Nc-1)!!}{(\Nc!)^2}
\end{equation}
(\ref{eq:FyodorovSommersLimiting1}) and (\ref{eq:FyodorovSommersLimiting2}) match exactly the limiting forms derived in Eqs.~165 and~166 of Ref.~\cite{FyoSom97}.

\subsection{Marginal distribution of the partial time delays for arbitrary symmetry class}
\label{sec:Partial}

We now consider the marginal distribution of the partial time delays for arbitrary symmetry class and show that it takes a rather simple form in the weak coupling limit.
We follow the formulation introduced by Gopar and Mello \cite{GopMel98} for $\Nc=1$ and further generalised in \cite{SavFyoSom01} for arbitrary $\Nc>1$, although these papers did not consider specifically the weak coupling limit.

When all channels are equally coupled, Eq.~\eref{eq:TheModel} implies a relation between the eigenvalues of the two scattering matrices
$ \EXP{\I\theta_a} = \big(\Sbar+\EXP{\I\theta_a^{(0)}}\big)\big(1+\Sbar^*\,\EXP{\I\theta_a^{(0)}}\big)^{-1}$. This leads to the following relation between the partial times $\tilde{\tau}_a=\partial_\varepsilon\theta_a$ and $\tilde{\tau}_a^{(0)}=\partial_\varepsilon\theta_a^{(0)}$ \cite{GopMel98,SavFyoSom01}:
\begin{equation}
  \tilde{\tau}_a 
   = f(\theta_{a}^{(0)})\,\tilde{\tau}_a^{(0)}
\end{equation}
where
\begin{equation}
   f(\theta)=\frac{1-|\Sbar|^2}{\big| 1+\Sbar^*\EXP{\I\theta}\big|^2}
   = \frac{ \frac{2\fss}{1-\fss^2} }{ \frac{1+\fss^2}{1-\fss^2} + \cos\theta }
   = \frac{1}{g+\sqrt{g^2-1}\,\cos\theta}
   \:.
\end{equation}
(We have used $\Sbar=(1-\fss)/(1+\fss)$, choosing $\fss\in[0,1]$).
We can therefore write the distribution as
$\mpart{\Nc}{\beta}(\tau)
= \smean{ \delta(\tau - f(\theta_{a}^{(0)})\,\tilde{\tau}_a^{(0)} }_{\theta_{a}^{(0)},\tilde{\tau}_a^{(0)}}
$.
Now we use the fact that for perfect coupling, the phase shifts are uniformly distributed and uncorrelated from the partial time delays. As a consequence~:
\begin{equation}
  \label{eq:GMandSFS}
  \mpart{\Nc}{\beta}(\tau)
  =
  \int_0^{2\pi}\frac{\D\theta}{2\pi}\,\frac{1}{f(\theta)}\,
  \mpart{\Nc}{\beta}^{(0)}(\tau/f(\theta))
  \:,
\end{equation}
where \cite{SavFyoSom01}~:
\begin{equation}
  \label{eq:SavinFyodorovSommers2001}
  \mpart{\Nc}{\beta}^{(0)}(\tau)
  =\frac{1}{\Nc}\sum_{a=1}^\Nc \mean{ \delta( \tau - \tilde{\tau}_a^{(0)} ) }
  =\frac{(\beta/2)^{1+\beta\Nc/2}}{\Gamma(1+\beta\Nc/2)}\,
  \frac{\EXP{-\beta/(2\tau)}}{\tau^{2+\beta\Nc/2}}
\end{equation}
The representation \eref{eq:GMandSFS}, written under a slightly different form in \cite{SavFyoSom01}, generalizes the one obtained by Gopar and Mello for $\Nc=1$ \cite{GopMel98}.
We can make this integral representation more explicit through the rescaling
\begin{equation}
  \Rmpart{\Nc}{\beta}(\rt)
  = \frac{\beta}{4\sqrt{g^2-1}}\:
  \mpart{\Nc}{\beta}\left(\tau=\frac{\beta\,\rt}{4\sqrt{g^2-1}} \right)
\end{equation}
with $2\sqrt{g^2-1}=1/\fss-\fss$.
Some algebra gives the form
\begin{eqnarray}
  \label{eq:MarginalPartialForallBeta}
  &
  \hspace{-2cm}
  \Rmpart{\Nc}{\beta}(\rt)
  =
  \frac{1}{\Gamma(1+\frac{\beta\Nc}{2})}\,\rt^{-2-\beta\Nc/2}
  \\\nonumber
  & \times
  \int_{0}^{\pi}\frac{\D\theta}{\pi}\,
  \left(\frac{2\sqrt{g^2-1}}{g+\sqrt{g^2-1}\,\cos\theta}\right)^{1+\beta\Nc/2}\,
  \exp\left\{-\frac{2\sqrt{g^2-1}}{\rt\,(g+\sqrt{g^2-1}\,\cos\theta)}\right\}
  \:.
\end{eqnarray}
Before taking the limit of weak coupling ($g\to\infty$), we find more convenient to change the variable as $u=\tan^2(\theta/2)$, leading to the exact expression
\begin{eqnarray}
  \label{eq:MarginalPartialTimesGeneralForm}
  &
  \hspace{-2cm}
  \Rmpart{\Nc}{\beta}(\rt)
  =
  \frac{1}{\pi\,\Gamma(1+\frac{\beta\Nc}{2})} \, \rt^{-2-\beta\Nc/2}
  \\\nonumber
  &\times
  \int_0^\infty\frac{\D u}{\sqrt{u}}\,(1+u)^{\beta\Nc/2}\,
  \left(\frac{ 1-\fss^2 }{ 1+\fss^2\,u }\right)^{1+\beta\Nc/2}\,
  \exp\left\{
    -\frac{1+u}{\rt}\,\frac{1-\fss^2}{1+\fss^2\,u}
  \right\}
  \:.
\end{eqnarray}
It will be also convenient to express the cumulative distribution
\begin{equation}
  \label{eq:CumulativeMarginalPartial}
  \int_{\rt}^\infty\D y\,\Rmpart{\Nc}{\beta}(y)
  =\frac{1}{\pi}
  \int_0^\infty\frac{\D u}{\sqrt{u}(1+u)}\,
  \frac{\gamma\left(1+\frac{\beta\Nc}{2},\frac{1+u}{\rt}\,\frac{1-\fss^2}{1+\fss^2\,u}\right)}
       {\Gamma(1+\frac{\beta\Nc}{2})}
\end{equation}
where $\gamma(a,z)$ is the incomplete Gamma function~\cite{gragra}.

\subsubsection{Limit $\fss\to0$.}

The integral representation \eref{eq:MarginalPartialTimesGeneralForm} is the most appropriate in order to study the limit of weak coupling $\fss\to0$. It makes clear that the distribution takes the simple form in this limit~:
\begin{eqnarray}
  \label{eq:MarginalPartialTimeGoparMelloMethod}
  \lim_{\fss\to0}\Rmpart{\Nc}{\beta}(\rt)
  = \frac{1}{\sqrt{\pi}\,\Gamma(1+\beta\Nc/2)}
  \frac{\EXP{-1/\rt}}{\rt^{2+\beta\Nc/2}}\,
  U\left(  \frac{1}{2} , \frac{\beta\Nc+3}{2} , \frac{1}{\rt} \right)
  \:.
\end{eqnarray}
where $U(a,c,z)$ is the Kummer function~\cite{AbrSte64}.
It is quite remarkable to obtain a universal form describing the full distribution in this limit.

The expression further simplifies in the unitary case ($\beta=2$) as the Kummer function can be expressed as a sum
\begin{eqnarray}
  \hspace{-2.5cm}
  U\left(  \frac{1}{2} ,\Nc+ \frac{3}{2} , \frac{1}{\rt} \right)
  = \frac{1}{\sqrt{\pi}}
  \int_0^\infty\frac{\D u}{\sqrt{u}}(1+u)^{\Nc}\EXP{-u/\rt}
  = \frac{1}{\sqrt{\pi}}\sum_{n=0}^\Nc C_\Nc^n\,\Gamma(n+1/2)\,\rt^{n+1/2}
  \:.
\end{eqnarray}
Using $\Gamma(n+1/2)=2^{-n}(2n-1)!!\sqrt{\pi}$, we get
\begin{equation}
  \label{eq:MPartialLimitBeta2}
    \lim_{\fss\to0}
  \Rmpart{\Nc}{2}(\rt)
    = \frac{\EXP{-1/\rt}}{\sqrt{\pi}\,\rt^{3/2}}\,
    \sum_{n=0}^\Nc  \frac{(2n-1)!!}{n!\,(N-n)!\,2^n} \,\rt^{-\Nc+n}
    \:.
\end{equation}
which is in exact correspondence with \eref{eq:MarginalPartialTimesWC}, as it should, although the two derivations are quite different.

\subsubsection{Far tail ($\rt\gg1/\fss^2$).}

A more careful analysis of the integral \eref{eq:MarginalPartialTimesGeneralForm} shows that for small but finite $\fss$, the distribution presents a different behaviour for $t\gg1/\fss^2$.
In this case, noticing that
$  (1-\fss^2)(1+u)/\big[t\,(1+\fss^2u)\big] \ll 1$,
we can replace the exponential in \eref{eq:MarginalPartialTimesGeneralForm} by unity, which shows that the distribution has a power law tail with large exponent
$\Rmpart{\Nc}{\beta}(\rt)\simeq(a_\Nc/g^3)(\rt/g^2)^{-2-\beta\Nc/2}$, with $2\fss\simeq1/g$ and where the coefficient $a_\Nc$ can be easily found and is given below.


\subsection{Marginal distribution of the proper time delays in the unitary case}
\label{sec:Proper}

\begin{figure}[!ht]
\centering
\includegraphics[width=0.45\textwidth]{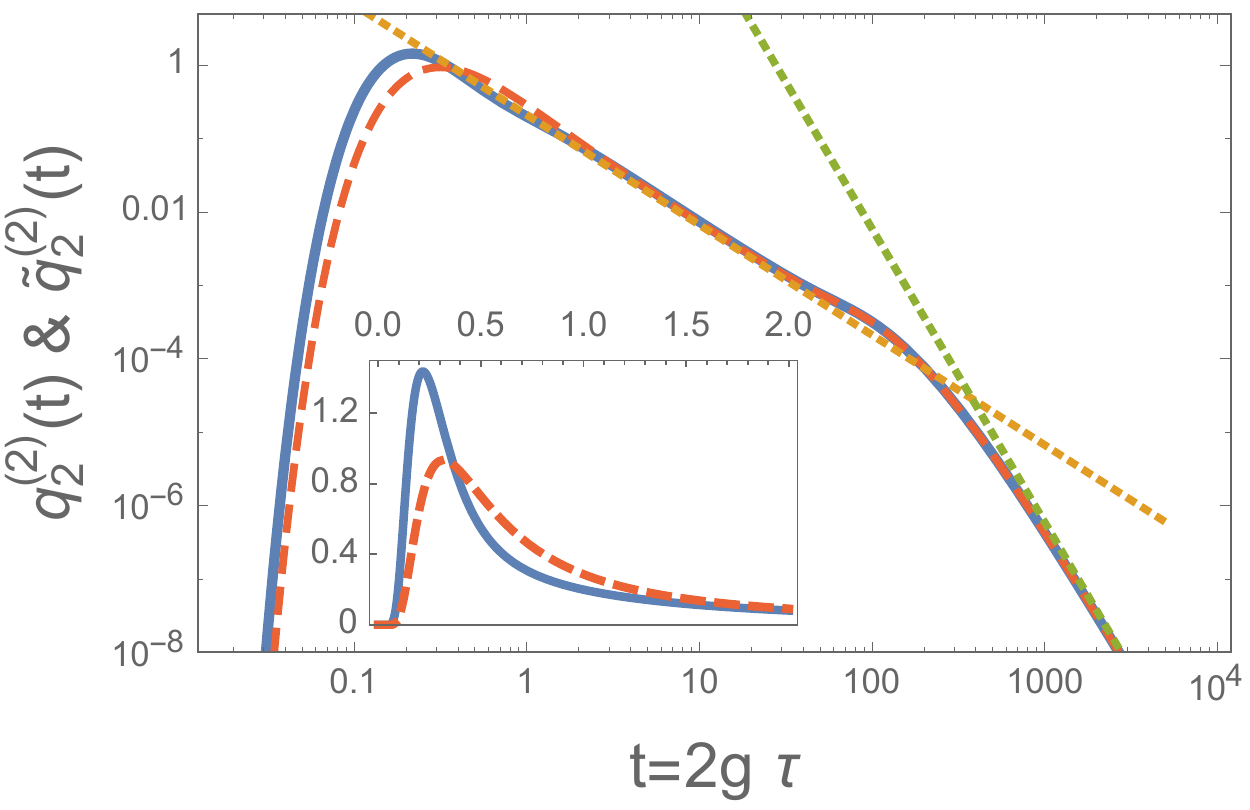}
\hfill
\includegraphics[width=0.45\textwidth]{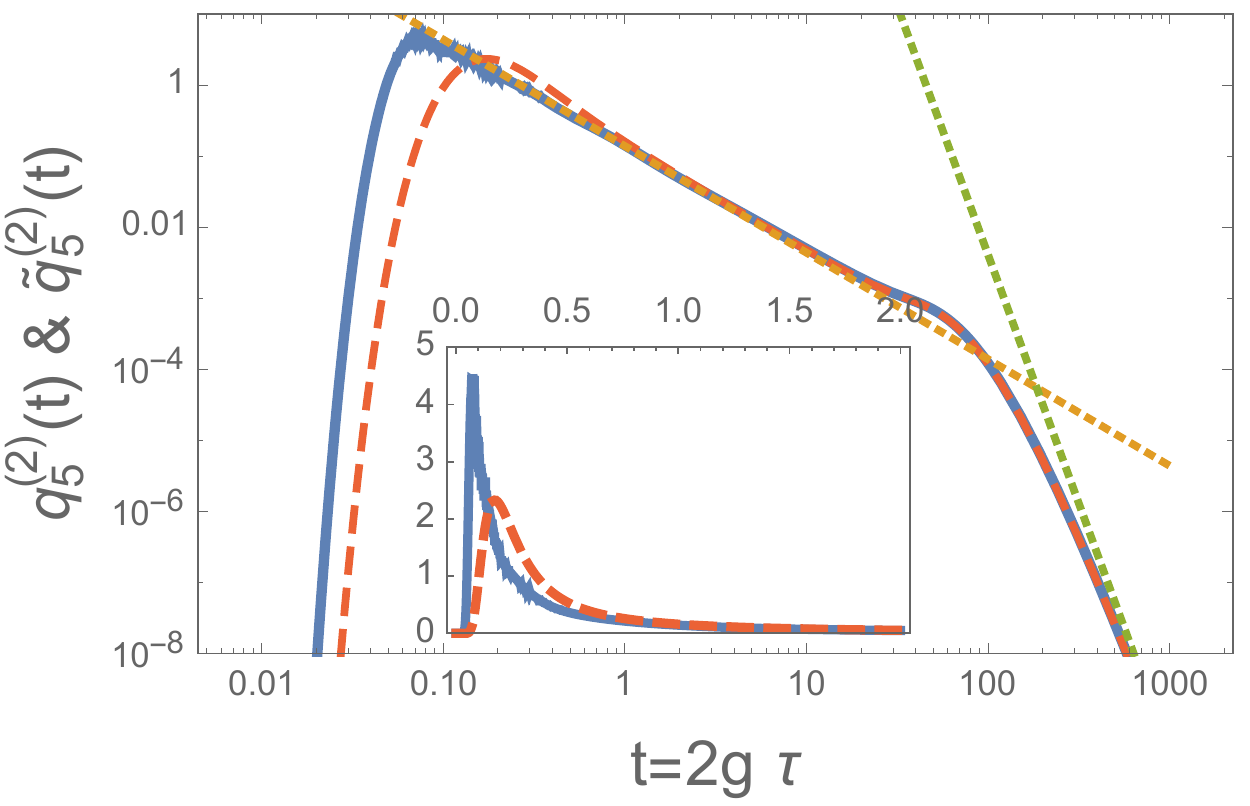}
\caption{\it Comparison between the marginal distributions for proper times (continuous blue line) and partial times (dashed red line) for $\Nc=2$ and $\Nc=5$.
Coupling parameter is $g=2/\coupl-1=10$.
The dotted lines are the two power laws with exponents $3/2$ and $2+\Nc$.
  }
  \label{fig:ComparisonProperPartial}
\end{figure}

The exact explicit form for the marginal distribution of the proper time delays was only found for the unitary case in Ref.~\cite{SomSavSok01}~:
\begin{equation}
  \label{eq:SommersSavinSokolov2001a}
  \mprop{\Nc}{2}(\tau)
  = \frac{1}{\Nc\tau}\sum_{n=0}^{\Nc-1}
  \left(
    F_n\derivp{B_n}{\tau} - B_n\derivp{F_n}{\tau}
  \right)
\end{equation}
where
\begin{eqnarray}
  \label{eq:SommersSavinSokolov2001b}
  B_n &= \frac{1}{n!} \left(-\derivp{}{g}\right)^n
  \left[
    I_0(\sqrt{g^2-1}/\tau)\, \EXP{-g/\tau}
  \right]
  \\
  \label{eq:SommersSavinSokolov2001c}
  F_n &= \sum_{m=0}^n\frac{1}{(2m+1)!}
  \left(\derivp{^2}{g^2}-\frac{2}{\tau}\derivp{}{g}\right)^m g^n
  \:.
\end{eqnarray}
This analytic solution is still quite complicate. Already for $\Nc=5$, the plot with the software {\tt Mathematica} shows some irregularities (cf. Fig.~\ref{fig:ComparisonProperPartial}).
The above explicit expressions become of limited use for plotting the distribution at larger $\Nc$. (Note, however, that one can alternatively use an integral representation of the exact distribution that can be inferred from the analysis of \cite{SomSavSok01}).
Hence it is instructive to extract limiting behaviours directly from Eq.~\eref{eq:SommersSavinSokolov2001a}. Let us now study this point.

The expression \eref{eq:SommersSavinSokolov2001a} can be rewritten in terms of the scaling variable as
\begin{equation}
    \label{eq:SommersSavinSokolov2001aBIS}
  \Rmprop{\Nc}{2}(\rt)
  = \frac{2g}{\Nc\rt}\sum_{n=0}^{\Nc-1}
  \left(
    F_n\derivp{B_n}{\rt} - B_n\derivp{F_n}{\rt}
  \right)
  \:.
\end{equation}
We now discuss the structure of the functions $B_n$ and $F_n$.
The functions $F_n$ can be computed systematically from \eref{eq:SommersSavinSokolov2001c}~:
\begin{eqnarray}
  F_0 &= 1 \\
  F_1 &= g \, \left( 1 - \frac{2}{3\rt} \right) \\
  F_2 &= g^2 \, \left( 1 - \frac{4}{3\rt}+\frac{4}{15\rt^2} \right) + \frac{1}{3} \\
  F_3 &= g^3 \, \left( 1 - \frac{2}{\rt}+\frac{4}{5\rt^2} - \frac{8}{105\rt^3} \right)
       + g \, \left( 1 - \frac{2}{5\rt} \right) \\
      & \vdots \hspace{2cm} \vdots
\end{eqnarray}
For the following, it is sufficient to identify the first and last terms in the contribution of order $g^n$~:
\begin{eqnarray}
  F_n = g^n\,\left(
     1 - \frac{(\cdots)}{\rt} + \cdots + (-1)^n\frac{2^{2n}n!}{(2n+1)!\,\rt^n}
  \right)
  + g^{n-2} \, (\cdots) + \cdots
\end{eqnarray}
Note that the term $t^{-n}$ corresponds to $(-2/\tau)^n\partial_g^ng^n$.

We now focus on the functions $B_n$'s in the large $g$ limit and restrict ourselves to the regime $\tau\ll g$, i.e. $t\ll g^2$.
In this case we can write
\begin{equation}
  B_0 \simeq \frac{1}{2g} \psi(\rt=2g\tau)\,,
  \hspace{1cm}\mbox{with }
  \psi(\rt) \eqdef \sqrt{\frac{\rt}{\pi}}\EXP{-1/\rt}
\end{equation}
which considerably simplifies Eq.~\eref{eq:SommersSavinSokolov2001b}
\begin{equation}
  B_n \simeq \frac{1}{2g^{n+1}}
  \sum_{m=0}^n \frac{(-\rt)^m}{m!}\,\psi^{(m)}(\rt)
  \:.
\end{equation}
We now remark that the calculation of the derivatives $\psi^{(m)}(\rt)$ can be simplified in the two limiting cases $\rt\gg1$ or $\rt\ll1$.

\paragraph{Limit $\rt\gg1$ (and $\rt\ll g^2$).}

For large $\rt$, the derivatives of $\psi(\rt)$ are dominated by derivation of the power law $\sqrt{\rt}$ in $\psi(\rt)$, hence
\begin{equation}
 B_n \simeq \alpha_n\frac{\psi(\rt)}{2g^{n+1}}
 \hspace{1cm}\mbox{with }
  \alpha_n\eqdef\sum_{m=0}^n \frac{(-1)^m(\frac12-m+1)_m}{m!}
\end{equation}
Then the distribution is dominated by the term
\begin{equation}
  \Rmprop{\Nc}{2}(\rt)
  \simeq \frac{2g}{\Nc\rt}\sum_{n=0}^{\Nc-1}
    F_n\derivp{B_n}{\rt}
    \simeq \frac{\psi(\rt)}{2\Nc\rt^2} \sum_{n=0}^{\Nc-1}\alpha_n
    \:.
\end{equation}
Finally we can write
\begin{equation}
  \label{eq:C35}
  \Rmprop{\Nc}{2}(\rt)
  \simeq b_\Nc \, \rt^{-3/2}
\end{equation}
where
\begin{equation}
 b_\Nc = \frac{1}{2\sqrt{\pi}\,\Nc}
   \sum_{n=0}^{\Nc-1} (\Nc-n) \frac{(-1)^n(\frac12-n+1)_n}{n!}
   = \frac{(2\Nc-1)!!}{\sqrt{\pi}\,\Nc!2^\Nc}
   \:.
\end{equation}
This is precisely the coefficient of the marginal for partial times, Eq.~\eref{eq:CoeffBn} for $\beta=2$. We have also checked that it coincides with the precise behaviour given in Ref.~\cite{SomSavSok01} for large $\Nc$
\begin{equation}
  \mprop{\Nc}{\beta}(\tau)
  \simeq \frac{1}{\pi\sqrt{2\Nc g}} \tau^{-3/2}\qquad
  \mbox{ for }
  1/g\ll \tau \ll g
  \:.
\end{equation}

\paragraph{Limit $\rt\ll1$.}

The derivatives of $\psi(\rt)$ are dominated by derivation of the exponential, hence
\begin{equation}
 B_n \simeq \frac{\psi(\rt)}{2g^{n+1}}
  \sum_{m=0}^n \frac{(-1)^m}{m!\rt^m}
  \simeq \frac{\psi(\rt)}{2g^{n+1}}\frac{(-1)^n}{n!\rt^n}
\end{equation}
Using the expansion of $F_n$, we get
\begin{equation}
  \Rmprop{\Nc}{2}(\rt)
  \simeq \frac{2g}{\Nc\rt}
    F_{\Nc-1}\derivp{B_{\Nc-1}}{\rt}
    \simeq
    \frac{2^{2(\Nc-1)}}{\sqrt{\pi}\,\Nc(2\Nc-1)!}\,\frac{\EXP{-1/\rt}}{\rt^{2N+1/2}}
    \:.
\end{equation}
This behaviour is different from the one obtained for partial times, cf.~Eq.~\eref{eq:MarginalPartialTimesWC}.

\subsection{Comparison of the two marginal distributions}

Although the two distributions $\mpart{\Nc}{2}(\tau)$ and $\mprop{\Nc}{2}(\tau)$ look at first sight quite different (see plots in linear scale in Fig.~\ref{fig:ComparisonProperPartial}), we have showed that they precisely coincide as soon as $\rt\gg1$~:
not only the power law $\rt^{-3/2}$ coincide, but also the precise coefficient.
We interpret this as a manifestation of the fact that, for $\rt\gg1$, the two distributions are dominated by isolated resonances.
Although we have not extracted from (\ref{eq:SommersSavinSokolov2001a},\ref{eq:SommersSavinSokolov2001b},\ref{eq:SommersSavinSokolov2001c}) the behaviour for $\rt\gg g^2$, based on the isolated resonance picture, we assume that the distributions also coincide in this regime as well. We write
$\Rmprop{\Nc}{\beta}(\rt)\simeq\Rmpart{\Nc}{\beta}(\rt)$ for $\rt\gg1$, i.e.
\begin{equation}
  \mprop{\Nc}{\beta}(\tau) \simeq \mpart{\Nc}{\beta}(\tau)
  \hspace{1cm}\mbox{for }
  \tau \gtrsim \tlow
\end{equation}
where $\tlow\sim1/g\sim\fss$, as long as resonances can be considered as isolated, according to the discussion of the introduction (see Fig.~\ref{fig:ComparisonProperPartial}).
The dependence of the cutoff in the channel number is determined below. Hence this is a strong difference between the weak coupling and perfect coupling regimes: while the two marginals strongly differ in the latter, the almost coincide in the former
(See Fig.~\ref{fig:SketchPC}).

\subsubsection{Crossovers.}

Before summarizing the different limiting behaviours, we determine the precise value where the distribution crosses over from one limiting behaviour to another in the limit of large $\Nc$.

The asymptotic form of the coefficients will be useful (we only consider the unitary case)~:
\begin{equation}
  a_\Nc \simeq \sqrt{2}\,\left(\frac{4}{\Nc}\right)^\Nc \frac{\EXP{\Nc}}{\pi\Nc}
  \:,\hspace{1cm}
  b_\Nc \simeq \frac{1}{\pi\sqrt{\Nc}}
  \:,\hspace{1cm}
  c_\Nc \simeq \frac{\EXP{2\Nc}}{4\pi\,\Nc^{2\Nc+1/2}}
  \:.
\end{equation}

Let us denote $\tup$ the crossover position between the two last limiting behaviours~: we write
$
  b_\Nc \, (g/\tup)^{3/2} = a_\Nc\, (g/\tup)^{2+\Nc}
$.
Using the asymptotics of the coefficients, one gets the upper cutoff (in unit of $\Ht$)
\begin{equation}
  \label{eq:DefUpperCutoffProper}
  \tup \simeq 4\,\mathrm{e}\, \frac{g}{\Nc}
\end{equation}

Similarly, we determine the position where the distribuition crosses over between the universal $\tau^{-3/2}$  power law and the $\tau\to0$ behaviour.
As the two distributions $\Rmprop{\Nc}{2}(\rt)$ and $\Rmpart{\Nc}{2}(\rt)$ differ in this regime, we have to discuss separately the cases of partial and proper times.
We consider first the case of partial times~:
we write
$
  \tilde{c}_\Nc\,\rt^{-\Nc-3/2}\,\EXP{-1/\rt} = b_\Nc\, \rt^{-3/2}
$
leading to the equation
$
  {1}/{\rt} + \Nc \,\ln t  = \Nc - \Nc \,\ln\Nc -(1/2)\ln2
$.
Thus we obtain the lower cutoff $\tilde\rt_\mathrm{lower}\simeq1/\Nc$, i.e.
\begin{equation}
  \tilde\tlow \simeq \frac{1}{\Nc g}
  \:.
\end{equation}
For the proper time we write
$
  c_\Nc\,\rt^{-2\Nc-1/2}\,\EXP{-1/\rt} = b_\Nc\, \rt^{-3/2}
$,
leading to the equation
$
  {1}/{\rt} + 2\Nc \,\ln t  = 2\Nc - 2\Nc \,\ln\Nc -2\ln2
$,
i.e. $\rt_\mathrm{lower}\simeq1/(2\Nc)$.
The cutoff for the proper time is half the cutoff for the partial times
\begin{equation}
  \label{eq:DefLowerCutoffProper}
  \tlow \simeq \frac{1}{2\Nc g} \simeq \frac{1}{2}\,\tilde\tlow
  \:.
\end{equation}

\subsubsection{Summary of the limiting behaviours.}

In conclusion, we have seen that the marginal distribution presents three limiting behaviours~:
\begin{equation}
\label{eq:LimitsMarginalPartialTimes}
\hspace{-2cm}
  \Rmpart{\Nc}{\beta}(\rt)
  \simeq
  \frac{1}{2g} \mpart{\Nc}{\beta}\left(\tau\simeq\frac{t}{2g}\right)
  \underset{g\to\infty}{\simeq}
  \left\{
    \begin{array}{ll}
    \displaystyle
    \tilde{c}_\Nc\,\rt^{-\beta\Nc/2-3/2}\,\EXP{-1/\rt}
    & \mbox{for }     \rt\lesssim 1/\Nc
    \\[0.25cm]
    \displaystyle
    b_\Nc \,\rt^{-3/2}
    & \mbox{for }    1/\Nc \lesssim\rt \lesssim g^2/\Nc
    \\[0.25cm]
    \displaystyle
    \frac{a_\Nc}{g^3}\,\left(\frac{g^2}{\rt}\right)^{2+\beta\Nc/2}
    & \mbox{for }   \rt\gtrsim  g^2/\Nc
    \end{array}
  \right.
\end{equation}
where the three coefficients are
\begin{eqnarray}
  \label{eq:CoeffCn}
  \tilde{c}_\Nc &= \frac{1}{\sqrt{\pi}\,\Gamma(1+\beta\Nc/2)}
  \:,
  \\
  \label{eq:CoeffBn}
  b_\Nc &= \frac{1}{\pi}\, \frac{\Gamma(1/2+\beta\Nc/2)}{\Gamma(1+\beta\Nc/2)}
  \:,
  \\
  \label{eq:CoeffAn}
  a_\Nc &= \frac{2^{1+\beta\Nc}}{\sqrt\pi}\, \frac{\Gamma(1/2+\beta\Nc/2)}{\Gamma(1+\beta\Nc/2)^2}
  \:.
\end{eqnarray}

The marginal distribution of the proper times is only known in the unitary case~:
\begin{equation}
  \label{eq:LimitsMarginalProperTimes}
\hspace{-2cm}
  \Rmprop{\Nc}{2}(\rt)
  \simeq
  \frac{1}{2g} \mprop{\Nc}{2}\left(\tau\simeq\frac{t}{2g}\right)
  \underset{g\to\infty}{\simeq}
  \left\{
    \begin{array}{ll}
    \displaystyle
    c_\Nc\,\rt^{-2\Nc-1/2}\,\EXP{-1/\rt}
    & \mbox{for } \rt\lesssim 1/\Nc
    \\[0.25cm]
    \displaystyle
    b_\Nc  \,\rt^{-3/2}
    & \mbox{for }  1/\Nc \lesssim\rt \lesssim g^2/\Nc
    \\[0.25cm]
    \displaystyle
    \frac{a_\Nc}{g^3} \left(\frac{g^2}{\rt}\right)^{\Nc+2}
    & \mbox{for } \rt\gtrsim  g^2/\Nc
    \end{array}
  \right.
\end{equation}
where the coefficient obtained above is
\begin{equation}
   c_\Nc = \frac{2^{2(\Nc-1)}}{\sqrt{\pi}\,\Nc(2\Nc-1)!}
   \:.
\end{equation}

s

\subsubsection{Moments of partial times and proper times.}

We have recalled in the introduction the variance of the partial and proper times.
In particular, in the unitary case, we have seen that the second moment is
$\mean{\tilde{\tau}_a^2}\simeq g/\Nc^2$ for weak coupling $g\gg1$ (compared to $\simeq1/\Nc^3$ for perfect coupling $g=1$).
We now analyse more into detail the moments of the partial times and of the proper times in the weak coupling limit in the unitary case.

\paragraph{Positive moments.}

In the weak coupling regime, the disributions $\mprop{\Nc}{\beta}(\tau)$ and $\mpart{\Nc}{\beta}(\tau)$ coincide for $\tau\gg1/g$, i.e. the part of the distributions which controls the \textit{positive} moments~:
\begin{equation}
  \smean{\tau_a^k}\simeq\smean{\tilde\tau_a^k}
  \hspace{1cm}\mbox{for }  k<1+\beta\Nc/2
\end{equation}
($\smean{\tilde\tau_a^k}=\smean{\tau_a^k}=\infty$ for $k\geq1+\beta\Nc/2$).

The calculation of the moments is dominated by the $\tau^{-3/2}$ tail, cutoff by the faster decay $\tau^{-2-\beta\Nc/2}$ above $\tup$, where the cutoff was determined above.
We can estimate the positive moments as
\begin{equation}
  \mean{\tau_a^k} \simeq \int^{\tup}\D\tau\,\frac{b_\Nc}{\sqrt{g}}\,\tau^{k-3/2}
  \sim  \frac{b_\Nc}{\sqrt{g}}\,\tup^{k-1/2}
\end{equation}
leading to the typical scale
\begin{equation}
       \mean{\tau_a^k}^{1/k} \sim \frac{1}{\Nc^{1/k}}\, \tup^{1-1/k}
       \sim \frac{g^{1-1/k}}{\Nc}
       \sim \frac{1}{\Nc\,\coupl^{1-1/k}}
\end{equation}
for $k<1+\beta\Nc/2$.

\paragraph{Negative moments.}

The \textit{negative} moments are controlled by the lower cutoff introduced above.
We can write
\begin{equation}
  \mean{\tau_a^{-k}} \simeq \int_{\tlow}\D\tau\,\frac{b_\Nc}{\sqrt{g}}\,\tau^{-k-3/2}
  \sim  \frac{b_\Nc}{\sqrt{g}}\,\tlow^{-k-1/2} \sim (\Nc g)^k
\end{equation}
i.e.
\begin{equation}
       \mean{\tau_a^{-k}}^{-1/k} \simeq 2^{-1-1/(2k)} \mean{\tilde\tau_a^{-k}}^{-1/k}
       \sim \tlow
       \sim \frac{1}{\Nc g}
       \sim \frac{\coupl}{\Nc}
  \:.
\end{equation}

\end{appendix}


\section*{References}

%

\end{document}